\documentclass[lettersize,journal]{IEEEtran}
\usepackage{amsmath,amsfonts}
\usepackage{algorithmic}
\usepackage{algorithm}
\usepackage{array}
\usepackage[caption=false,font=normalsize,labelfont=sf,textfont=sf]{subfig}
\usepackage{textcomp}
\usepackage{stfloats}
\usepackage{url}
\usepackage{verbatim}
\usepackage{graphicx}
\usepackage{cite}
\usepackage{subfig}                 
\usepackage{overpic}      
\usepackage{multirow}
\usepackage{makecell}
\usepackage{bm}         
\usepackage{lipsum}
\usepackage{tabularx}
\makeatletter
\renewcommand{\maketag@@@}[1]{\hbox{\m@th\normalsize\normalfont#1}}%
\makeatother
\begin{document}

\title{Cost-Effective RF Fingerprinting Based on \\ Hybrid CVNN-RF Classifier with Automated \\ Multi-Dimensional Early-Exit Strategy}

\author{Jiayan Gan, Zhixing Du, Qiang Li, Huaizong Shao, Jingran Lin, Ye Pan, Zhongyi Wen and Shafei Wang
\thanks{This research was supported by the National Natural Science Foundation of China under Grant 62171110.
Note that this is an extension of our previous work \cite{ref1} presented at the 2023 IEEE International Conference on Communications.
(Corresponding author: Qiang Li.)}
\thanks{J. Gan, Z. Du and Z. Wen are with the University of Electronic Science and Technology of China, Chengdu 611731, China and the Nanhu Laboratory, Jiaxing 314000, China (e-mail: \{jiayan\_gan, zhixingdu, 2019151201001\}@std.uestc.edu.cn)} %
\thanks{Q. Li, H. Shao, J. Lin, Y. Pan and S. Wang are with the University of Electronic Science and Technology of China, Chengdu 611731, China and the Laboratory of Electromagnetic Space Cognition and Intelligent Control, Beijing 100089, China (e-mail: \{lq, hzshao, jingranlin, pany\}@uestc.edu.cn).}
}


\IEEEpubid{0000--0000/00\$00.00~\copyright~2024 IEEE} 

\maketitle

\begin{abstract}
While the Internet of Things (IoT) technology is booming and offers huge opportunities for information exchange, 
it also faces unprecedented security challenges. 
As an important complement to the physical layer security technologies for IoT, 
radio frequency fingerprinting (RFF) is of great interest due to its difficulty in counterfeiting. 
Recently, many machine learning (ML)-based RFF algorithms have emerged. In particular, deep learning (DL) has shown great benefits in automatically extracting complex and subtle features from raw data with high classification accuracy. 
However, DL algorithms face the computational cost problem as the difficulty of the RFF task and the size of the DNN have increased dramatically.
To address the above challenge, this paper proposes a novel cost-effective early-exit neural network consisting of a complex-valued neural network (CVNN) backbone with multiple random forest branches, called hybrid CVNN-RF.
Unlike conventional studies that use a single fixed DL model to process all RF samples, our hybrid CVNN-RF considers differences in the recognition difficulty of RF samples and introduces an early-exit mechanism to dynamically process the samples.
When processing ``easy'' samples that can be well classified with high confidence, the hybrid CVNN-RF can end early at the random forest branch to reduce computational cost. 
Conversely, subsequent network layers will be activated to ensure accuracy. To further improve the early-exit rate, 
an automated multi-dimensional early-exit strategy is proposed to achieve scheduling control from multiple dimensions within the network depth and classification category.
Finally, our experiments on the public ADS-B dataset show that the proposed algorithm can reduce the computational cost by 83{\%} while improving the accuracy by 1.6{\%} under a classification task with 100 categories.
\end{abstract}


\section{Introduction}
\IEEEPARstart{W}{ith} the development of communication technology, there will be more than 50 billion Internet of Things (IoT) devices by 2025, according to IDC forecast \cite{ref2}. 
These connected devices can easily exchange information with each other, but their identities are highly susceptible to being duplicated and impersonated by hackers, 
posing significant security problems for IoT. How to accurately identify and authenticate these IoT devices is the primary issue that needs to be addressed for device management. 
In addition to traditional cryptography-based authentication, radio frequency fingerprinting (RFF) is one of the most promising means to enhance physical layer security. 
Since it can identify wireless devices by extracting the electromagnetic signal characteristics reflected by radio frequency (RF) circuit defects as hard-to-replicate fingerprints, 
RFF plays an important role in the field of wireless communication and IoT security \cite{ref3,ref4,ref5,ref6,ref7,ref8,ref9}.

\begin{figure}[!t]
  \centering
  \includegraphics[width=1.0\columnwidth]{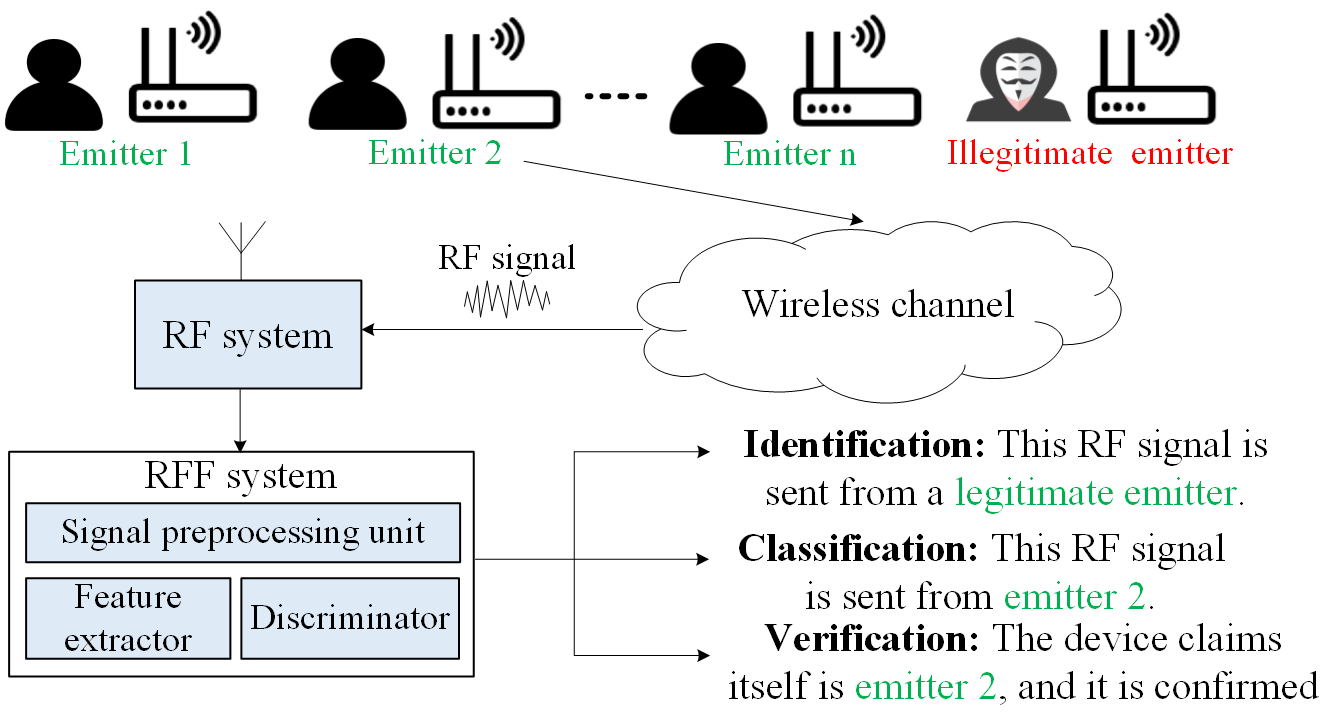}
  \caption{The application scenario of RFF.}
  \label{fig_1}
\end{figure}

For different wireless communication devices, the RF signals transmitted are slightly different due to minor manufacturing defects in analog components within the RF circuit including power amplifiers, 
digital-to-analog converters, frequency mixers, and band-pass filters \cite{ref10}. 
Therefore, RFF can exploit such signal difference to extract RF fingerprint which reflects the inherent characteristics of devices and is difficult to forge, 
even if they are from the same type or manufacturer. Typically, the application scenario of RFF is shown in Fig.1, involving emitters, an RF system, and an RFF system. The RF system can receive RF signals transmitted over wireless channels from emitters such as cell phones, radios, unmanned aerial vehicles, and other wireless communication devices. After that, the RFF system performs signal preprocessing, feature extraction, and discrimination on the received RF signals to accomplish three types of tasks, including identification, classification, or 
\IEEEpubidadjcol
verification. Identification is to determine whether the identity of an emitter is legitimate or not. 
Classification is to discriminate the specific identity (index/label) of the emitter under a closed set of all legitimate devices. Verification is to confirm whether the packet comes from the identity claimed by the emitter. All three tasks are important for IoT security, but this paper focuses more on the closed-set classification task in the RFF domain.

In recent years, many RFF methods have emerged, which can be divided into traditional machine learning (ML)-based methods and deep learning (DL)-based methods. Traditional ML-based methods usually require the incorporation of expert knowledge to manually extract RF signal features from the time domain or transform domain. Following this feature engineering, common machine learning classifiers can be used for the recognition and classification of RF signals. For example, Deng et al \cite{ref11} proposed an RFF method based on support vector machine (SVM) for broadband radio. During the classification of 3 AKDS700 radios, multidimensional permutation entropy features were used as inputs to the SVM. The final experiments demonstrated that the proposed method achieved 90\% classification accuracy. In \cite{ref12}, an RFF method based on random forest for ZigBee device was proposed by Patel et al. For feature engineering, multiple sets of non-parametric features including mode, median, mean, and linear model coefficients were generated and fed to the random forest. As in their experiments, 4 Texas Instruments ZigBee CC2420 devices were used for classification and 97\% accuracy was achieved on the random forest. Besides, Rehman et al. \cite{ref13} proposed a k-nearest neighbor (KNN)-based RFF method for Bluetooth device. To classify 7 built-in Bluetooth transceivers of smartphones, a KNN classifier with 3 nearest neighbors was employed for analyzing features obtained from the energy envelope of Bluetooth signals. The experimental results showed that their method achieved 99.9\% classification accuracy. For different types of RF signals, the above traditional ML-based RFF methods employ different combinations of hand-craft features and machine learning classifiers. These methods have achieved satisfactory classification performance on their own small-scale datasets. However, the manual feature engineering scheme lacks generality to a certain extent, and the applicability of these traditional ML-based methods under large-scale datasets remains to be examined.

Unlike traditional ML-based methods, DL-based methods can automatically extract high-level abstract features from input data without manual feature engineering. Due to their powerful learning and processing capabilities, many researchers have paid great attention to DL-based methods and designed deep neural networks (DNNs) specifically for RFF. In \cite{ref14}, Sankhe et al. proposed an RFF method named ORACLE for 802.11 device classification. They used a LeNet-like DNN with 5 layers in ORACLE, which consists of 2 convolution (CONV) layers and 3 fully connected (FC) layers. Under static ideal channel conditions, the experimental results showed that ORACLE achieved a classification accuracy of 98.6\% for 16 X310 USRP devices. Based on the research of ORACLE, Sankhe et al. \cite{ref15} further experimented on a real-world RF dataset, including 500 to 50 consumer-of-the-shelf (COTS) WiFi devices. For this dataset, an AlexNet-like DNN with deeper layers was used to replace the original LeNet-like DNN, where the number of CONV layers was increased to 8 and the number of FC layers was kept constant. The final experiments revealed that the classification accuracy ranged from 30\% to 85\% on datasets with different numbers of devices. In \cite{ref16}, an RFF method based on VGG-like DNN was proposed by Zong et al. Deeper than the AlexNet-like DNN, the VGG-like DNN consists of 14 CONV layers and 3 FC layers. In addition, batch normalization and dropout operations are added after each convolutional layer and the first two fully connected layers, where the batch normalization operation helps to accelerate the convergence of the DNN training and the dropout helps to reduce overfitting. Finally, their method achieves 99.7\% classification accuracy on 5 transmitters. Similar to the work in \cite{ref15}, Gritsenko et al. \cite{ref17} conducted a study on a real-world dataset with 500 ADS-B and 500 WiFi RF data. To handle this large-scale dataset, an AlexNet-like DNN and a ResNet-like DNN were used for experiments. Unlike the AlexNet-like DNN, the ResNet-like DNN has a deeper network depth with 50 layers and residual connectivity. As an important feature of ResNet-like DNN, residual connectivity \cite{ref18} can alleviate the problem of gradient vanishing and facilitate the training of DNN with a large number of layers. The final experiments showed that the accuracy performance of ResNet-like DNN outperforms that of the AlexNet-like DNN by 15\% and 7\% under the WiFi dataset and the ADS-B dataset, respectively. In addition to residual design, attention mechanism embedding and complex-valued model modification are also new and effective techniques for DNN designs. Gu et al. \cite{ref19} proposed a DAConv-ResNet for 802.11 device classification. With ResNet as the backbone, DAConv-ResNet introduces an attention module called dual-attention convolutional module, which integrates spatial attention mechanism and channel attention mechanism. By making the convolutional operation focus more on the important parts of the input signal, this module helps to enhance the representation learning capability of the DNN. From the experimental results on 10 X310 USRP devices, their method improves the classification accuracy by 1.5\% compared with the ResNet baseline. Besides, Yang et al \cite{ref20} employed a complex-valued neural network for drone classification. Different from traditional real-valued neural networks, the complex-valued neural network (CVNN) uses operators in complex-valued form instead of real-valued form to extracted subtle features from RF signals with in-phase and orthogonal parts. The experimental results show that their method achieves 99.5\% classification accuracy for 4 types of drone activity data, which is 1.85\% higher than the best real-valued neural network used in their work. In summary, most of the above DL-based methods are usually aimed at adjusting the model structure of DL algorithms to improve the classification accuracy by deepening the network and broadening the operator functionality, but rarely discuss the computational cost. While large DNNs tend to have better classification performance compared to small DNNs especially on difficult datasets, the computational cost of DNNs cannot be ignored as the model size increases. If the computational cost is too high, it is not friendly for real-time and power-constrained scenarios.


When dealing with real-world RF datasets, we have noticed that RF samples with different signal quality pose different recognition challenges. Regrettably, existing studies typically use the same algorithm to process all RF samples without fully considering the differences in sample recognition difficulty. What affects the classification accuracy is dependent on the indistinguishable part of the samples, which we identify as ``hard'' samples. Most samples are ``easy'' samples and can be well classified using the simplified algorithm without requiring the full computational effort. Based on the above, we propose to use a novel early-exit neural network to mitigate the computational cost problem, which ensures the accuracy of ``hard'' samples while processing ``easy'' samples at low cost. Specifically, our work makes the following contributions:

\begin{itemize}
  \item{To the best of our knowledge, this is the first attempt to use an early-exit neural network for RF signal classification, especially for RFF. Meanwhile, we experimentally verify its feasibility and advantages using a real-world RF dataset which consists of 100 wireless devices. }
  \item{We introduce a novel cost-effective CVNN-RF classifier that employs a complex-valued neural network (CVNN) as the backbone structure and multiple random forest branches as auxiliary classifiers. The final experiment reveals that random forest branches as auxiliary classifiers provide substantial benefits in terms of accuracy and computational cost compared to other classifiers..}
  \item{To further improve the early-exit rate and reduce the computational cost of the algorithm, we propose an automated multi-dimensional early-exit strategy, which includes automated early-exit condition generation, multi-branch early-exit, and multi-category early-exit.}
\end{itemize}

A preliminary conference version of this work has been presented in \cite{ref1}, where a simple 4-category radio classification task was verified. In this journal version, we have considered a more challenging 100-category classification task and upgraded the network model to complex CVNN. Moreover, the newly developed automated early-exit condition generation and multi-category early-exit strategy improve the early-exit rate in [1], thus further reducing the computational complexity. 

\section{Proposed Method}
As the scale of DNN gradually increases, the non-negligible computational cost and burden imposed by the deep and intensive computation of DNN limits the deployment on resource-limited hardware. After DNN training, the depth of DNN generally remains static and fixed, which means that samples of whatever classification difficulty will consume the full computational cost on the entire DNN. Manifestly, such static computation is not efficient enough. A new dynamic computation is desired to reduce the redundant computation of ``easy'' samples by adjusting the network depth in real time. 

\begin{figure}[!ht]
  \centering
  \includegraphics[width=1.0\columnwidth]{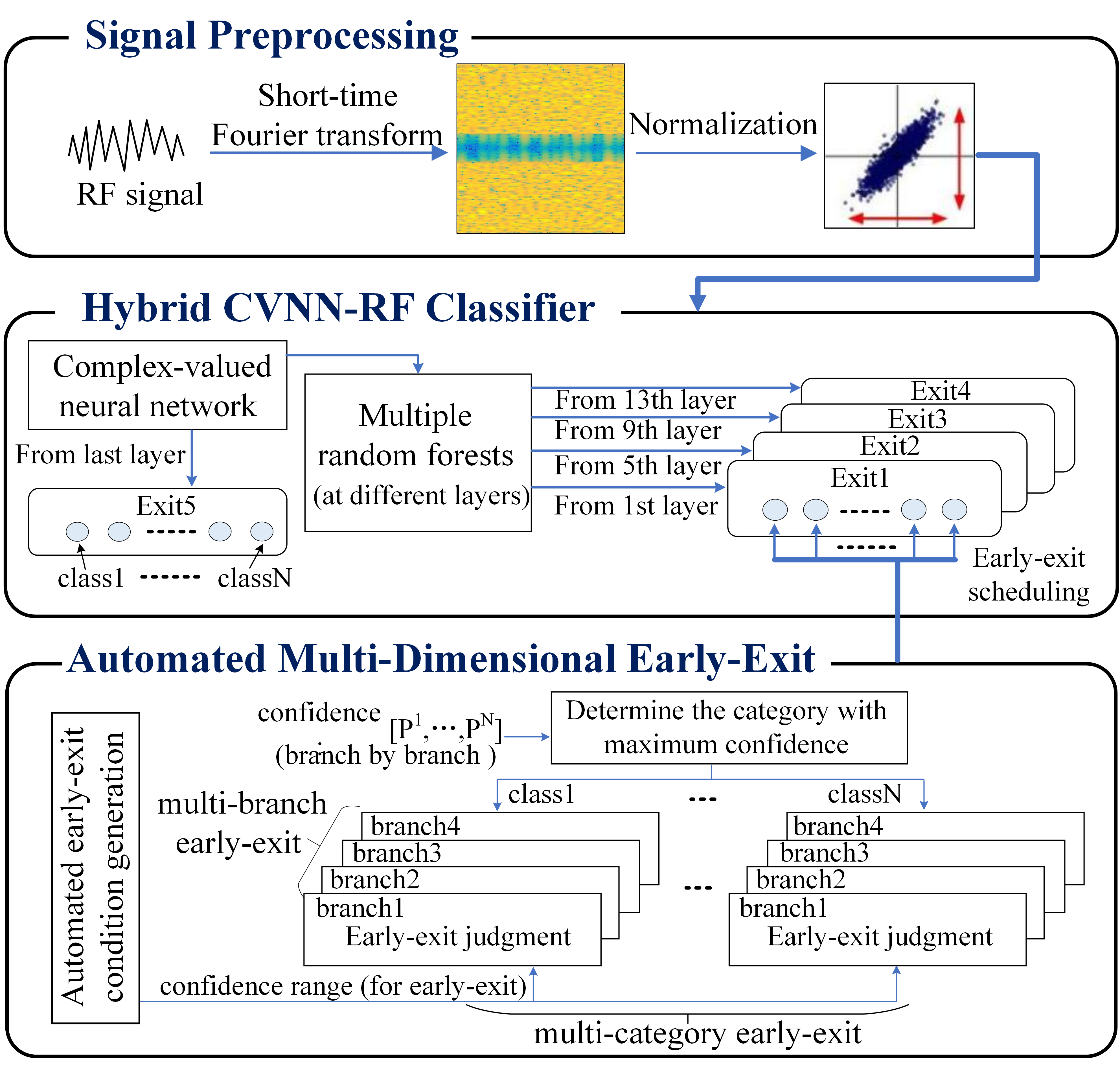}
  \caption{The overall framework of our proposed RFF method.}
  \label{fig_2}
\end{figure}

Drawing inspiration from the above considerations, we proposed our RFF method as illustrated in Fig. 2. Specifically, the method contains three parts: signal preprocessing, hybrid CVNN-RF classifier and automated multi-dimensional early-exit. In the signal preprocessing, we perform short-time Fourier transform (STFT) and normalization on the raw RF data. Automated multi-dimensional early-exit is used to control how early-exit is performed under different categories and branches. A hybrid CVNN-RF classifier is utilized to classify the different RF samples ranging with different classification difficulty and allow these samples to exit from different exit points. For those ``easy'' samples, they can exit from shallow exit points to reduce the computational cost. For those ``hard'' samples, deeper network layers will be inferred to obtain more complex and high-level features, ensuring high accuracy.

\subsection{Signal Preprocessing}

Neural network training can be challenging and may result in reduced performance when the raw data is directly fed into the network. 
Researchers typically preprocess RF signals to improve the accuracy of their algorithms. 
One widely used preprocessing method is the short-time Fourier transform (STFT) \cite{ref21}, 
which maps a one-dimensional time-domain signal to a joint distribution of time and frequency, 
preserving both time-domain and frequency-domain features of the signal. 
In this study, we utilized the STFT to preprocess RF signals and extract their time-domain and frequency-domain features.
The standard STFT output  $F_{stft}$ with length $N$ can be represented in the time and frequency dimensions as:

\begin{equation}
\label{equ1}
F_{stft}(t,f) = \sum_{n=-\infty}^{\infty} f[n] {\omega} [n-st] e^{-i\frac{2{\pi}n}{N} f}
\end{equation}
where $f$ is the original signal, $s$ is the stride, and $\omega$ is the window function.

Before feeding STFT data into a neural network, 
it is necessary to normalize the data to scale each sample to unit norm. Normalization \cite{ref22} ensures that all samples have the same range and facilitates the convergence of neural network training. In this study, we used min-max normalization to scale the STFT data to the range of {[-1, 1]}. 

\begin{equation}
\label{equ2}
X_{min-max} = \frac{2(x-x_{min})}{x_{max}-x_{min}} - 1
\end{equation}
where $x$ is the input data, ${X_{min-max}}$ is the output quantized data, and ${x_{min}}$ and ${x_{max}}$ are the minimum and maximum values of $x$, respectively.

\subsection{Hybrid CVNN-RF Classifier}
The design of our hybrid CVNN-RF classifier originated from the early-exit neural network. A typical early-exit neural network is usually composed of a backbone neural network and a branch classifier located at the hidden layer of the backbone. 
\begin{figure*}[!hb]
  \centering
  \includegraphics[width=2.0\columnwidth]{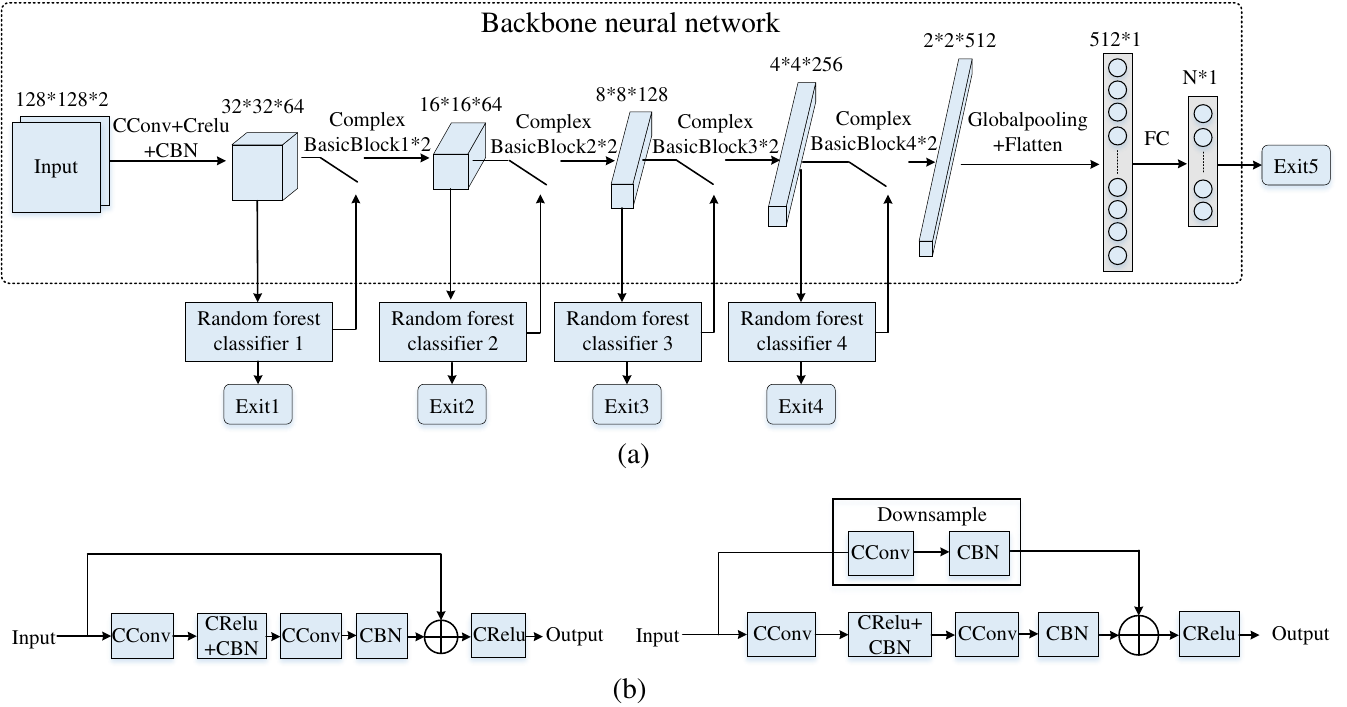}
  \caption{Proposed hybrid CVNN-RF classifier (a) the overall structure of CVNN-RF (b) the two structures of complex BasicBlock.}
  \label{fig_3}
\end{figure*}
Traditionally \cite{ref23,ref24,ref25}, the branch classifier is usually another low-cost and lightweight neural network branch including the CONV branch and the FC branch. 
In the process of inferring the early-exit neural network, confidence can be used as an important evaluation criterion for early-exit, which refers to the classification probability given by the branch classifier for each category. With the confidence, a simple and conventional early-exit strategy is to compare it with a preset threshold. 
If the confidence is higher than the preset threshold, the current sample is regarded as an ``easy'' sample that can exit early and the predicted label of the branch classifier will be output directly. 
Otherwise, the current sample is regarded as a ``hard'' sample and its hidden features need to be returned to the backbone neural network for further processing with the subsequent layers.

As one of the early-exit neural networks, our proposed hybrid CVNN-RF classifier uses CVNN as the backbone neural network and four random forest classifiers as auxiliary branch classifiers. 
Fig. 3(a) shows its overall structure. The backbone CVNN, called CResNet-18, is a ResNet-like DNN consisting of a complex convolutional (CCONV) layer, 
a complex relu (Crelu) activation layer, a complex batch normalization (CBN)layer, 
8 complex BasicBlocks, a global pooling layer, a flatten layer, and a FC layer. 
Among various DL models, CVNN \cite{ref26} is a novel model that exhibits powerful processing and learning capabilities in the field of RFF. However, a CCONV operation usually requires four times more computation than a standard CONV operation, which makes CVNN suffer from a high complexity problem. Our choice of CVNN as the backbone is mainly due to the expectation that our early-exit neural network can get significant recognition performance and cost benefits on CVNN. 
The input of CResNet-18 consists of two IQ channels of the STFT time-frequency map with a shape of 128×128. 
The output of CResNet-18 is activated with softmax function and consists of $N$ neurons, where $N$ denotes the number of categories under different classification tasks. 
Each intermediate feature map between the input and the output is represented by a cube marked with its dimension. Note that each cube is obtained by concatenating the real and imaginary parts of the complex operation results in channel dimension. 

\begin{figure}[!ht]
  \centering
  \includegraphics[width=1.0\columnwidth]{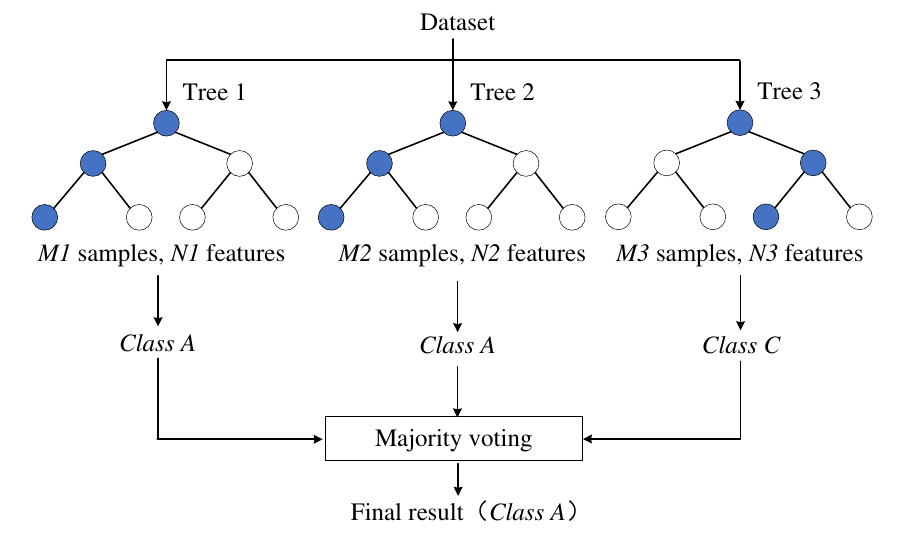}
  \caption{An example of random forest classifier.}
  \label{fig_4}
\end{figure}

Fig. 3(b) shows the two detailed structures of complex BasicBlock, 
where the first structure corresponds to the case with a stride size of 1 and the second structure corresponds to the case with a stride size of 2. 
Each complex BasicBlock is grouped in pairs, with the first using structure 2 and the other using structure 1. 
In total, there are four groups of complex BasicBlocks including complex BasicBlock1, complex BasicBlock2, 
complex BasicBlock3, and complex BasicBlock4 with channel sizes of 64, 128, 256, and 512. 
The first structure of complex BasicBlock consists of two CCONV layers, two Crelu layers, and two CBN layers. 
Between the input and the second CBN layer of BasicBlock, there is a direct residual path implemented by summation. 
Unlike the first structure, the second structure adds a downsample block in the residual path to avoid the dimensionality problem. 
Here the downsample block is implemented through a CCONV layer with stride of 2 and a CBN layer.

CCONV is the main operator of CVNN, derived from the complex extension of real-valued convolution. 
In CCONV, both the convolution kernel and the input signal are in complex form. 
Given a complex input of ${r = p+iq}$ (p and q are real matrices) and a complex convolution kernel of ${W = A+iB}$ ($A$ and $B$ are real matrices), the convolution result $v$ of the CCONV can be expressed as:

\begin{equation}
  \label{equ3}
  v = W \times  r = (A \times p) + i(B \times p + A \times q)
\end{equation}

It can be seen that the real part of $v$ is the difference between the real convolution $A\times p$ and $B \times q$, and the imaginary part of $v$ is the sum of the real convolution $B\times p$ and $A\times q$.

An alternative expression in matrix form is as follows:

\begin{equation}
  \label{equ4}
  \left[
  \begin{array}{c}
     Re(v) \\
     Im(v) \\
  \end{array}
  \right]
  =
  \left[
  \begin{array}{cc}
      A & -B \\
      B & A  \\
  \end{array}
  \right]
  \times
  \left[
  \begin{array}{c}
      p \\ 
      q \\
  \end{array}
  \right]
  \end{equation}
where $Re(v)$ and $Im(v)$ are the real and imaginary parts of $v$, respectively.

CBN involves scaling the hidden features to a standard normal complex distribution, which facilitates the convergence of CVNN training. The standard normalization for a complex input $h$ can be achieved by multiplying the inverse square root of the covariance matrix $C$  with the zero-centered data $(h-E[h])$:

\begin{equation}
  \label{equ5}
  {\tilde{h}} = (C)^{-0.5}(h-E[h])
\end{equation}
where,
\begin{normalsize}
\begin{equation}
  \begin{aligned}
  \label{equ6}
  C &= 
  \begin{pmatrix}
     C_{rr} & C_{ri} \\
     C_{ir} & C_{ii} \\
  \end{pmatrix}
  \\
  &=
  \begin{pmatrix}
      Cov(Re(h),Re(h)) & Cov(Re(h),Im(h)) \\
      Cov(Im(h),Re(h)) & Cov(Im(h),Im(h)) \\
  \end{pmatrix}
\end{aligned}
\end{equation}
\end{normalsize}

\leftline{where $Cov(\cdot)$ denotes the covariance function.} 

Similar to batch normalization for real values, CBN can also scale and shift the standard normalized data $\tilde{h}$ by setting the learnable parameters $\gamma$ and $\beta$ as follows:

\begin{equation}
  \label{equ7}
  CBN(h) = \gamma\tilde{h} + \beta
\end{equation}

Crelu activation introduces nonlinearity to CVNN and enhances its ability to learn more complex functional mapping relationships. For a complex input $z$, the $Relu()$ function is performed separately for its real and imaginary parts, as follows:
\begin{equation}
  \label{equ8}
  Crelu(z)=Relu(Re(z))+Relu(Im(z))i 
\end{equation}

In order to give the input samples more chances to exit early in CVNN, four random forest classifiers are set after the first CBN layer, the complex BasicBlock1, the complex BasicBlock2, and the complex BasicBlock3, respectively. Random forest \cite{ref27} is a popular machine learning algorithm that combines multiple decision trees for ensemble learning. Fig. 4 shows an example of a random forest classifier consisting of 3 decision trees. With bootstrap sampling, each decision tree is constructed from different subsets of the training samples and features. For example, the first decision tree uses $M1$ samples and $N1$ features, while the second decision tree uses $M2$ samples and $N2$ features. At each node of the decision tree, the feature with the best splitting effect is 
typically selected for decision based on Gini impurity or information gain. In the prediction phase of the random forest, the classification results $A$, $A$, and $C$ from 3 decision trees are aggregated by voting to obtain the final prediction result $A$. Owing to the random selection of samples and features as well as ensemble learning, the diversity of decision trees reduces the risk of overfitting and brings good generalization performance to random forest.
\subsection{Automated Multi-Dimensional Early-Exit}

To achieve good cost efficiency while ensuring accuracy, 
we propose an automated multi-dimensional early-exit strategy, 
including automated early-exit condition generation, multi-branch early-exit, and multi-category early-exit. 
For automated early-exit condition generation, it is responsible for automatically generating multiple early-exit confidence ranges $range_{exit}^{m,n}$ as the early-exit conditions, 
where $n$ ranges from 1 to the number of categories $N$ and $m$ ranges from 1 to the number of branches $M$. 
For multi-category early-exit, it performs different early-exit judgments according to different categories. 
Due to the different distribution and difficulty level of samples in each category, 
one early-exit condition cannot guarantee the individuality of the category. 
Thus, we consider using independent early-exit conditions for different categories after selecting the category with the highest confidence value. 
For multi-branch early-exit, it performs different early-exit judgments according to different branches. 
Compared to a single branch, multiple branches as exit points can achieve more flexible early-exit control. 
By performing early-exit judgments branch by branch, samples of different difficulty levels can exit early at different exit points.
During the early-exit judgment, the maximum prediction confidence of the branch classifier selected from $P^1, P^2,\dots,P^N$ is used to determine whether it falls within the confidence range generated by the automated early-exit condition generation. 
If it is, the neural network inference will end and exit early from this branch classifier. Otherwise, it will continue to the next layer. 
Unlike the conventional early-exit strategy that only uses a single threshold as early-exit condition, our multi-dimensional early-exit strategy ensures fine-grained early-exit control for each branch and each category, which has higher flexibility and efficiency.


\begin{algorithm}[!t]
  \begin{small}
  \caption{Automated Multi-dimensional early-exit strategy (deployment phase)}\label{alg:alg1}
  \begin{algorithmic}
  \STATE  \textbf{Require:}
  \STATE  \quad\text{The correct labels on validation data $L_{correct}$ } 
  \STATE  \quad\text{The predicted labels of backbone on validation data $L_{nn}$ } 
  \STATE  \quad\text{The predicted labels of branch on validation data $L_{branch}$} 
  \STATE  \quad\text{The classification confidence of branch on validation data $Pval$} 
  \STATE  \quad\text{The number of validation data $D$} 
  \STATE  \quad\text{The number of segments $S$} 
  \STATE  \quad\text{The accuracy tolerance $T$} 
  \STATE  \textbf{Ensure:}
  \STATE  \hspace{0.2cm} \text{The early-exit confidence ranges} $range_{exit}^{m,n}$
  \vspace{0.2cm}
  \hrule
  \vspace{0.2cm}
  \begin{footnotesize}
  \STATE  \textbf {Conduct the following step for each category $n$ and each branch $m$}
  \end{footnotesize}

  \STATE  \quad\textbf{Step1: confidence range generation}
  \STATE  \qquad $P_{val}^{m,n} = P_{val}^{m,n}.sort()$
  \STATE  \qquad \text{for $b$ in range($S$):}
  \STATE  \qquad \quad $r^b =[ P_{val}^{m,n} (b*int(D/S)), P_{val}^{m,n} ((b+1)*int(D/S))]$

  \begin{footnotesize}
  \STATE  \qquad \text{Generate confidence range $r^b$, where $b$ is the index of the segment}
  \end{footnotesize}

  \STATE  \quad\textbf{Step2: exit flag generation}
  \begin{footnotesize}
    \STATE  \qquad\text{Within the limited range $r^b$, pick out the label $LC_b$, $LN_b$,} 
    \STATE  \qquad\text{ $LB_b$ from the original label  $L_{correct}^n$, $L_{nn}^n$, $L_{branch}^{m,n}$, respectively}
  \end{footnotesize}
  \STATE  \qquad\text{for $b$ in range($S$):}
  \STATE  \qquad\quad $Acc_{branch}^b $ = $sklearn.metrics.accuracy\_score (LC_{b}, LB_{b})$
  \STATE  \qquad\quad $Acc_{nn}^b  = sklearn.metrics.accuracy\_score (LC_b, LN_b)$
  \STATE  \qquad\quad \text{if $Acc_{nn}^b-Acc_{branch}^b<T$: }
  \STATE  \qquad\qquad \text{$Exit^b$ = 1}
  \STATE  \qquad\quad \text{else:}
  \STATE  \qquad\qquad \text{$Exit^b$ = 0}
  \begin{footnotesize}
    \STATE  \qquad\text{Generate exit flag $Exit^b$, where $b$ is the index of the segment}
  \end{footnotesize}
  \STATE  \quad\textbf{Step3: confidence range merge for early-exit}
  \STATE  \qquad\text{for $b$ in range($S$):}
  \STATE  \quad\qquad\text{if $Exit^b$ = 1:}
  \STATE  \qquad\qquad\text{ $range_{exit}^{m,n}$ = $merge(range_{exit}^{m,n},r^b)$ }
  \STATE  \quad\qquad\text{else:}
  \STATE  \qquad\qquad\text{ continue }
  \STATE  \textbf{End}
  
  \end{algorithmic}
  \label{alg1}
  \end{small}
  \end{algorithm}
  
In more detail, our early-exit strategy is presented in the form of pseudo-code, 
as shown in Algorithm~\ref{alg1} and Algorithm~\ref{alg2}. It can be divided into a deployment phase and an implementation phase. 
Before the early-exit strategy deployment, the backbone neural network and branch classifier are trained using the original training data and the corresponding feature data at the early-exit layer. 
After training, the backbone neural network can infer on all validation data to obtain predicted labels $L_{nn}$. 
Similarly, the branch classifier can infer on all test data or validation data to obtain the classification confidence $P_{test}$ or $P_{val}$ with predicted labels $L_{branch}$. 
In addition to the above four parameters, the correct label of the validation data $L_{correct}$, 
the number of segments $S$, the number of validation data $D$, and the accuracy tolerance $T$ are also used as inputs to our early-exit strategy.

Algorithm~\ref{alg1} shows the pseudo-code of our early-exit strategy in the deployment phase. For each category $n$ and each branch $m$, we perform the following steps:

{\it{Step1: confidence range generation}}

The confidence $P_{val}^{m,n}$ in the range of $[0,1]$ are divided into $n$ segments in ascending order according to the equal number of samples in each segment. 
At the same time, multi-segment confidence ranges corresponding to the start and end points of each segment are generated and denoted as a plurality of variables $r^b$, 
where $b$ is the index of the segment.

{\it{Step2: exit flag generation}}

Within the restricted range $r^b$, filter from the original label $L_{correct}^n$, $L_{nn}^n$, $L_{branch}^{m,n}$ and obtain the label $LC_b$, $LN_b$, $LB_b$, respectively. 
Comparing with the correct labels $LC_b$, we count the percentage of correct prediction labels $LB_b$ and $LN_b$ of the branch classifier and backbone neural network, 
and obtain the corresponding accuracy $Acc_{branch}^b $ and $Acc_{nn}^b$ within the confidence range $r^b$. 
If the difference between the accuracy $Acc_{branch}^b$ and the accuracy $Acc_{nn}^b$ is within a certain accuracy tolerance $T$, 
we mark the exit flag $Exit^b$ of confidence range $r^b$, as 1, which means that the samples in confidence range $r^b$, need to exit early. Instead, the exit flag $Exit^b$ is marked as 0.

\begin{algorithm}[!t]
  \begin{small}
  \caption{Automated Multi-dimensional early-exit strategy (implementation phase)}\label{alg:alg2}
  \begin{algorithmic}
  \STATE  \textbf{Require:}
  \STATE  \quad\text{The classification confidence of branch on test data $P_{test}$ } 
  \STATE  \quad\text{The early-exit confidence ranges $range_{exit}^{m,n}$} 
  \STATE  \textbf{Ensure:}
  \STATE  \hspace{0.2cm} \text{The early-exit judgment result}
  \vspace{0.2cm}
  \hrule
  \vspace{0.2cm}
  \begin{footnotesize}
  \STATE  \textbf {Conduct the following step for each branch $m$}
  \end{footnotesize}

  \STATE  \quad\textbf{Step1: category determination }
  \STATE  \qquad $d = argmax_{n} $ ${P_{test}}$
  \STATE  \qquad \text{Determine the category $d$ with maximum confidence}

  \STATE  \quad\textbf{Step2: early-exit judgment}

  \STATE  \qquad\text{if $P_{test}^{m,d}$ $\in$ $range_{exit}^{m,d}$:}
  \STATE  \qquad\qquad\text{the test sample exits early }
  \STATE  \qquad\text{else:}
  \STATE  \qquad\qquad\text{the test sample goes to the next neural network layer} 
  \STATE  \textbf{End}
  
  \end{algorithmic}
  \label{alg2}
  \end{small}
  \end{algorithm}

{\it{Step3: confidence range merge for early-exit}}

To obtain the final early-exit confidence ranges $range_{exit}^{m,n}$, 
we merge those confidence ranges that can be used for early-exit, 
that is, the confidence ranges with the exit flag $Exit^b$ equal to 1 in step2.

Algorithm~\ref{alg2} shows the pseudo-code of our early-exit strategy in the implementation phase. 
For each branch $m$, we perform the following steps:

{\it{Step1: category determination}}

The category $d$ with the maximum confidence among all categories is selected for later early-exit judgment.

{\it{Step2:  early-exit judgment}}

If the confidence $P_{test}^{m,d}$ belongs to the confidence range $range_{exit}^{m,d}$ generated in the deployment phase, 
the test sample exits early and the predicted label of branch $m$ is used as the final identification result. 
Otherwise, the test sample at early-exit branch needs to be further processed by the rest of the neural network layers.

\begin{figure}[!ht]
  \centering
  \includegraphics[width=1.0\columnwidth]{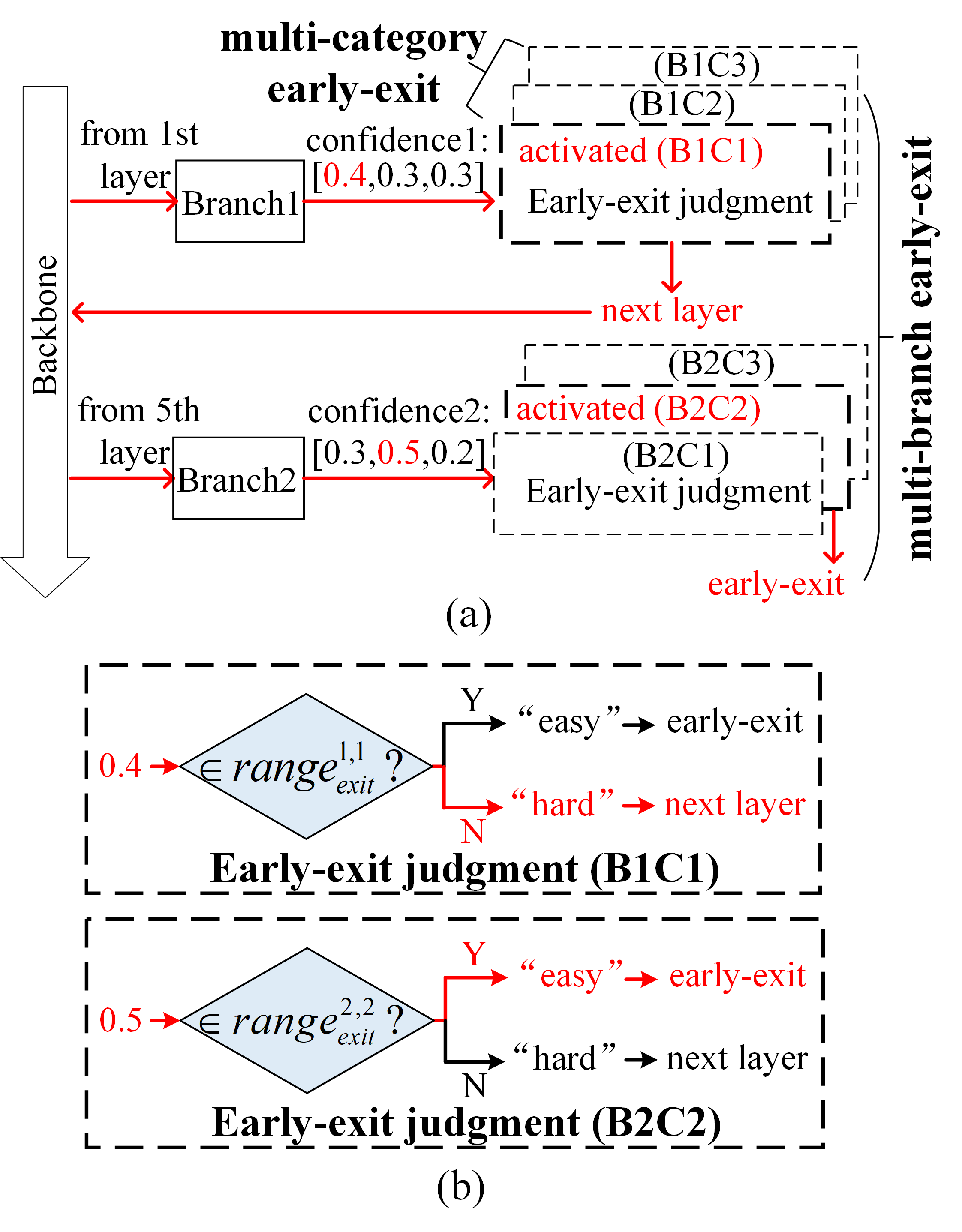}
  \caption{An example of using our early-exit strategy (with a category count of 3 and a branch count of 2) (a) the overall flow of our early-exit strategy (b) the details of early-exit judgment.}
  \label{fig_5}
\end{figure}

Fig. \ref{fig_5} shows an example of using our early-exit strategy, illustrated with 2 branches and 3 categories. The overall flow of the early-exit strategy is shown in Fig. 5(a). For branch 1 and branch 2, there are separate early-exit judgement function modules ($B1C1$, $B1C2$, $B1C3$, $B2C1$, $B2C2$ and $B2C3$) for each category. In this example, the branch 1 determines that the current test sample belongs to the category 1 based on the confidence vector of $[0.4, 0.3, 0.3]$ and gives the judgement result for returning to the next layer of the backbone. The branch 2 determines that the current test sample belongs to the category 2 based on the confidence vector of $[0.3, 0.5, 0.2]$ and gives the judgement result for early-exit. Fig. 5(b) gives the specific details of the early-exit judgement. For the branch 1, the confidence level 0.4 for the category 1 is the highest, so the early-exit judgement function module $B1C1$ is activated. Since 0.4 does not fall within the early-exit range $range_{exit}^{1,1}$, the current test sample is considered as a ``hard'' sample that need to be returned to the next layer of the backbone. For the branch 2, the confidence level of 0.5 for the category 2 is the highest, so the early-exit judgement function module $B2C2$ is activated. Since 0.5 falls within the early-exit range $range_{exit}^{2,2}$, the current test sample is considered as an ``easy'' sample that can exit early. Here the early-exit range $range_{exit}^{1,1}$ and $range_{exit}^{2,2}$ are determined in advance and can be referred to the deployment phase of our early exit strategy.
\begin{figure}[!ht]
  \centering
  \includegraphics[width=1.0\columnwidth]{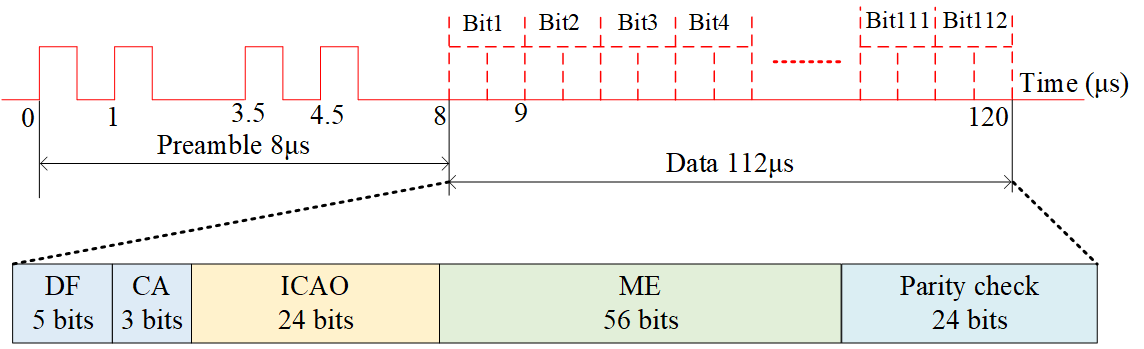}
  \caption{The frame format of a complete ADS-B signal.}
  \label{fig_6}
\end{figure}

\section{Experimental Results and Analysis}
\subsection{Experiment Setup}
\subsubsection{Dataset}
\
\newline 
\indent In this paper, a real-world public ADS-B dataset \cite{ref28} is used to validate the performance of our algorithm, 
which is collected at the Tianjin Civil Aviation industrialization base using the software-defined radio platform (SM200B). 
The ADS-B signal is a type of RF signal that is periodically broadcast in plain text form by commercial aircraft to help the air traffic control (ATC) center obtain its location and status information. 
For air traffic management and flight safety, the study of RFF for ADS-B signals is of great importance. 

Fig. 6 illustrates the frame format of a complete ADS-B signal. Its duration is typically 120$\mu$s, of which the first 8$\mu$s is the preamble segment and the last 112$\mu$s is the data segment. As the frame header of an ADS-B signal, the preamble segment consists of four pulses at the moments of 0$\mu$s, 1$\mu$s, 3.5$\mu$s, and 4.5$\mu$s, respectively. The data segment starts at the moments of 8$\mu$s and lasts until 120$\mu$s, which contains the 5-bit Downlink Format (DF) information, the 3-bit transponder capability (CA) information, the 24-bit International Civil Aviation Organization (ICAO) aircraft address information, the 56-bit message (ME) information, and the 24-bit parity identifier (PI) information. According to different message types, the ME field contains various information such as airborne position (longitude, latitude, altitude, etc.), airborne velocity, and aircraft operation status. These messages are not unique or fixed for an individual ADS-B target, so they are not sufficient to reveal the identity of the target. However, the ICAO code can serve as the unique identity of the target, which may affect the learned RF fingerprint features due to the memory effect of the DNN. Therefore, we removed the ICAO code from the ADS-B signals with 6000 sample points in the original dataset and obtained the processed ADS-B signals with 4800 sample points. 

Three categories were chosen to plot the first 1000 sample points of the ADS-B amplitude signal, as shown in Fig. 7. From the figure, it can be observed that there are slight differences between the signals of different ADS-B targets even in the fixed information (preamble segment, DF, and CA) of the first 800 points. Although the positions of the pulses in the fixed information are the same, the crests of the pulses as well as their rising and falling edges show different degrees of distortion. To extract the RF fingerprints associated with these subtle distortion characteristics, DNNs offer a promising solution with their powerful learning and automatic feature extraction capabilities. 

\begin{figure}[!ht]
  \centering
  \includegraphics[width=1.0\columnwidth]{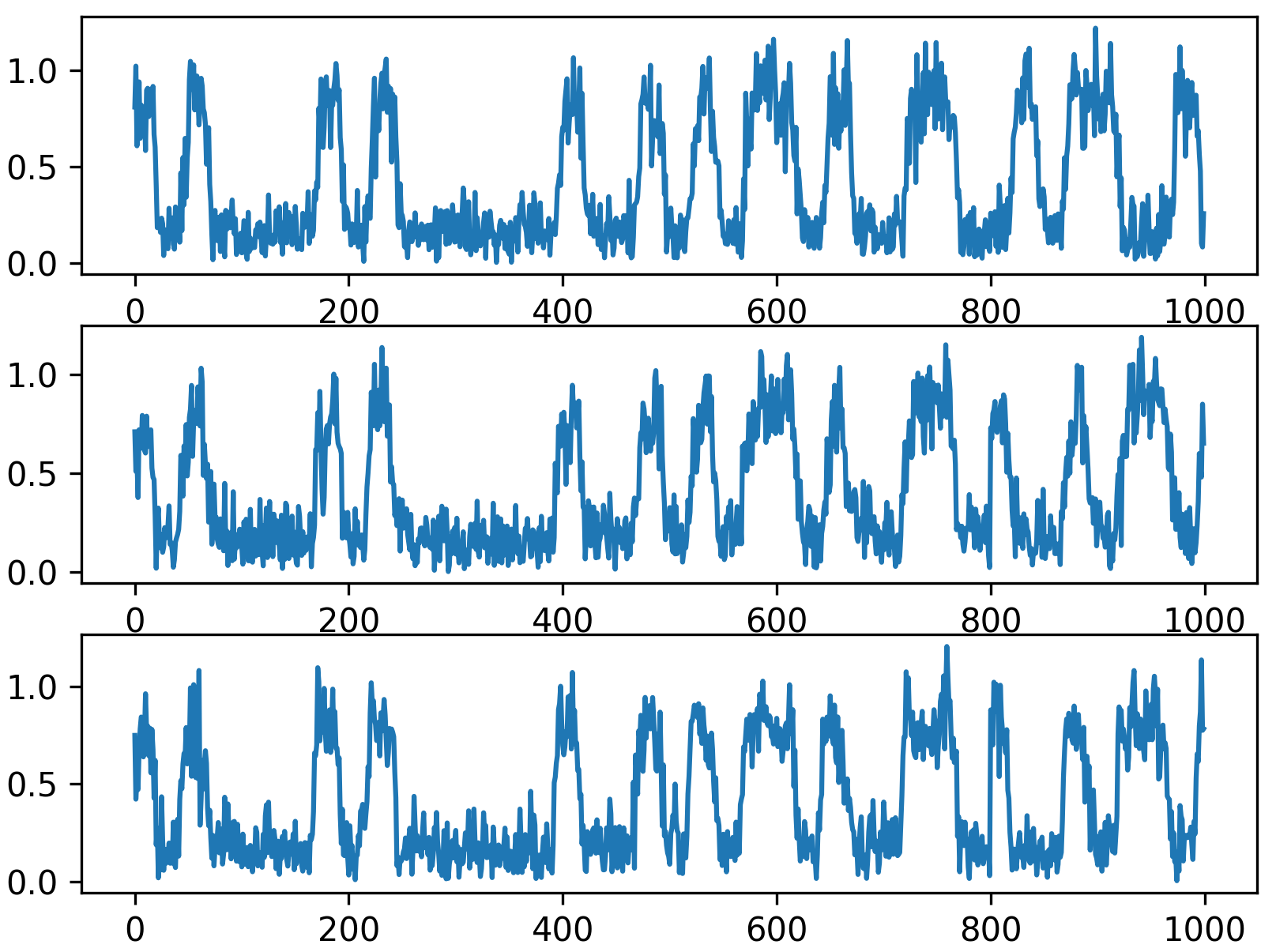}
  \caption{Amplitude signal waveforms for the first 1000 sample points of 3 ADS-B targets.}
  \label{fig_7}
\end{figure}

\begin{table}[!ht]
  \caption{\footnotesize {The hyperparameters in proposed method\label{tab:table1}}}
  \centering
  \setlength{\tabcolsep}{5.5pt}
  \renewcommand{\arraystretch}{2.0}
  \scalebox{1.0}{
  \begin{tabular}{c c c}
  \hline
  \hline
  \bf{Proposed method} & \bf{Hyperparameters} & \bf{Values}\\
  \hline
  \multirow{5}*{Hybrid CVNN-RF classifier} & Optimizer & 	Adam\\
  \cline{2-3}
  ~ & Epoch	& 300\\
  \cline{2-3}
  ~ & Batch size	& 1024\\
  \cline{2-3}
  ~ &Learning rate &	0.001 \\
  \cline{2-3}
  ~ & \makecell{Number and max depth \\of decision trees}	&400, 20\\
  \hline
  \multirow{2}*{\makecell{Automated multi-\\dimensional early-exit}} & Accuracy tolerance &	\makecell{10\% /7\% /5\%\\
  /2\% /0\% }\\ 
  \cline{2-3}
  ~ & \makecell{Number of entropy\\ range segments} & 15\\
  \hline
  Signal preprocessing	& \makecell{Window length of \\ STFT}	 & 128\\
  \hline
  \hline
  \end{tabular}
  }
\end{table}

For our experiments, we selected 10-100 categories of ADS-B signals,
where each category has 3600 samples. 
Considering the randomness of RF signal generation in the real world, 
we randomly assigned 60\% of the dataset as the training set, 30\% as the validation set and the remaining 10\% as the test set. In addition, 15 Monte Carlo experiments are conducted to ensure the reliability of the experimental results, and the median of the experimental results is taken as the most typical one for analysis.
\subsubsection{Target Platform and Framework}
\
\newline
\indent To ensure sufficient computational resources for our experiments, 
we chose an Intel Xeon Gold 6240 CPU and a NVIDIA Tesla V100S PCIe 32GB GPU as the target platform. 
Additionally, both keras and scikit-learn are utilized as basic frameworks for training the backbone neural network and the branch classifiers. 
\subsubsection{Experimental Setup Parameters}
\
\newline
\indent Table~\ref{tab:table1} shows several hyperparameters used in the hybrid CVNN-RF classifier,
the automated multi-dimensional early-exit strategy and the signal preprocessing. 
During the training process of the hybrid CVNN-RF classifier backbone, 
we utilized the Adam optimizer and set epoch to 300, batch size to 1024, 
and learning rate to 0.001. Meanwhile, the max depth and number of decision trees in all random forest branches are set to 20 and 400, respectively. 
Regarding the automated multi-dimensional early-exit strategy, the confidence range is divided into 15 segments and accuracy tolerance is chosen from 10\%, 7\%, 5\%, 2\%, and 0\%. 
As for the signal preprocessing, we set the window length of STFT to 128.
\begin{table}[!ht]
  \caption{\footnotesize {Accuracy (\%) comparison of the CVNN-RF backbone with other different algorithms \label{tab:table2}}}
  \centering
  \setlength{\tabcolsep}{17pt}
  \renewcommand{\arraystretch}{1.5}
  \scalebox{1.0}{
  \begin{tabular} {c c c c}
  \hline
  \hline
  \bf{Neural network}	& \bf{100}	& \bf{50}	& \bf{10}\\
  \hline
  CResNet-18 	&95.76	&96.32	& 99.11\\
  \hline
  ResNet-18	 & 92.12	&92.66	&98.80\\
  \hline
  ABL \cite{ref29}	 &90.67	  &91.23	&98.34\\
  \hline
  LightAMC \cite{ref30}	&84.01 	&85.16	    &95.49\\
  \hline
  MSCNN \cite{ref31}	 &85.07	  &86.12	&98.57\\
  \hline
  \hline
  \end{tabular}
  }
\end{table}
\begin{table}[!ht]
  \caption{\footnotesize {Comparison of the CVNN-RF backbone with other different algorithms in terms of parameters and computational costs\label{tab:table3}}}
  \centering
  \setlength{\tabcolsep}{17pt}
  \renewcommand{\arraystretch}{1.5}
  \scalebox{1.0}{
  \begin{tabular} {c c c c}
  \hline
  \hline
  \bf{Neural network}	& \bf{\makecell{Parameters \\ (MB)}}	& \bf{\makecell{Computational \\ costs (MFLOPs)}}\\
  \hline
  CResNet-18 	&5.66	&140.81\\
  \hline
  ResNet-18	 &2.83	&35.53\\
  \hline
  ABL \cite{ref29}	 &10.8	  &21.6\\
  \hline
  LightAMC \cite{ref30}	&0.32 	&23.48\\
  \hline
  MSCNN \cite{ref31}	 &1.89	  &55.11\\
  \hline
  \hline
  \end{tabular}
  }
\end{table}

\subsection{Performance of CVNN-RF Backbone}
In this section, we provide a comparison of the recognition performance between the CVNN-RF backbone CResNet-18 and its real-valued version ResNet-18. 
Other existing network structures in the field of RF signal recognition, including adaptive broad learning (ABL) \cite{ref29}, 
lightweight automatic modulation classification (LightAMC) \cite{ref30}, 
and multisampling convolutional neural network (MSCNN) \cite{ref31}, 
are also provided for comparison. These neural networks are reproduced under the same ADS-B dataset and their hyperparameters are set according to the corresponding  
article. Specifically, Table II shows the recognition performance of the above neural networks at 100, 50 and 10 categories. 
It can be seen that CResNet-18 can almost maintain higher recognition accuracy than the comparison method. 
As the number of classification categories increases, the performance of all methods declines to some degree. 
However, our CResNet-18 can still maintain high recognition accuracy. Taking a classification task with 100 categories as an example, 
the recognition accuracy of all compared methods drops below 95\%, while our CResNet-18 can still maintain 95.76\% recognition accuracy.

In terms of parameters and computational cost, we show the performance of each neural network as shown in Table III. While CResNet-18 achieves the highest accuracy, the complex-valued computation also imposes a high computational cost on CResNet-18, as previously mentioned in Section II-B. Compared to ResNet-18, the computational cost of CResNet-18 is about four times higher and the highest among all networks. Due to the complex-valued form of the network weights, the number of parameters in Cresnet-18 is also approximately twice that of ResNet-18 and is the second highest among all networks. Notably, our proposed CVNN-RF uses Cresnet-18 as the backbone and focuses more on reducing the computational cost associated with CVNN than the number of parameters. 
\begin{figure}[!ht]
  \centering
  \includegraphics[width=1.0\columnwidth]{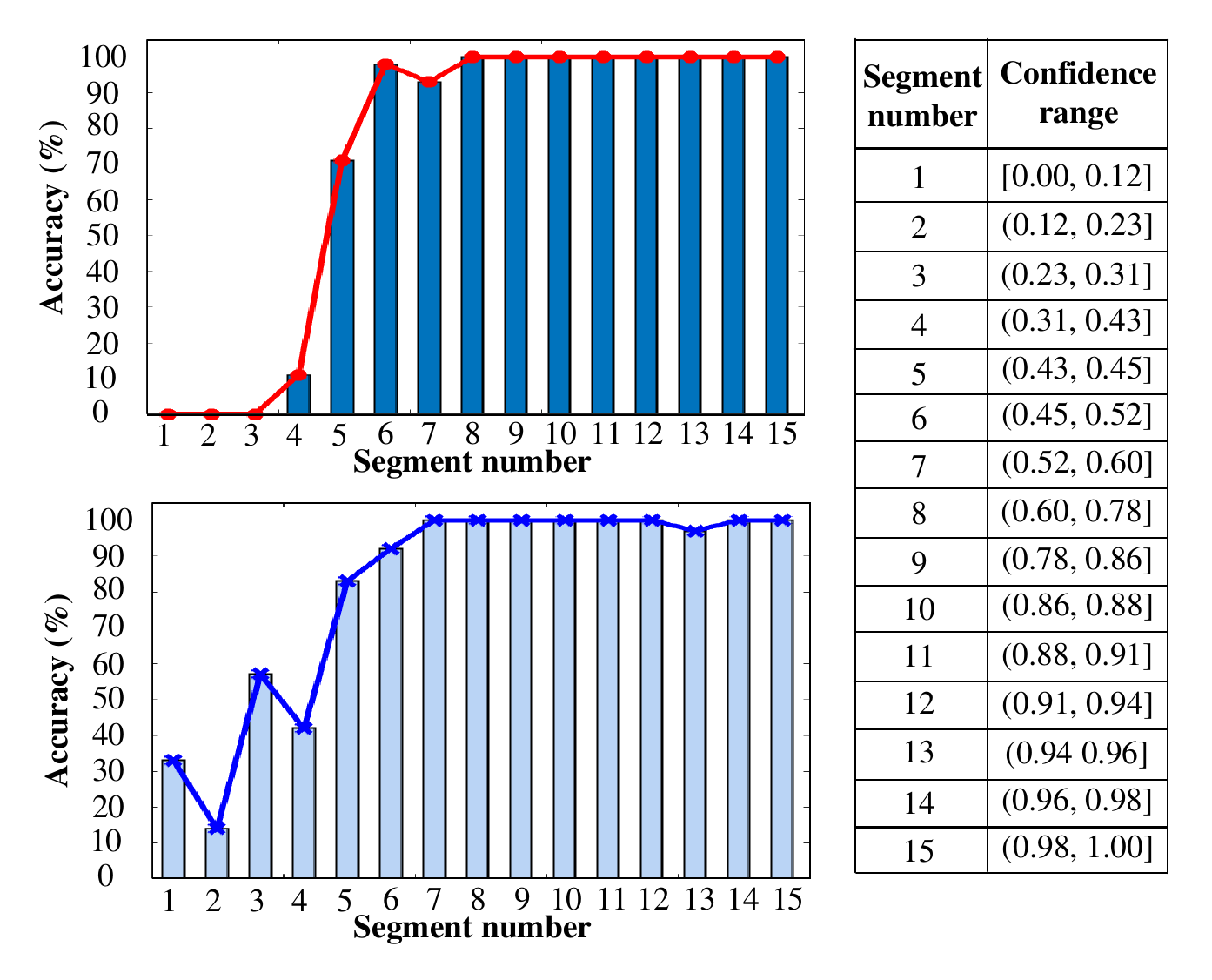}
  \caption{Statistical analysis of accuracy and confidence for the first branch and backbone under a selected category in the validation dataset.}
  \label{fig_8}
\end{figure}
With the introduction of our novel early-exit mechanism, we expect to realize significant computational cost reduction. For more analysis and experiments on CVNN-RF, please refer to the subsequent subsections.

\subsection{Statistical Analysis of Accuracy and Confidence for CVNN-RF}
In conventional early-exit neural networks, confidence is often used as a criterion to evaluate the sample difficulty. 
Samples with high classification confidence tend to be ``easy'' samples with high accuracy and samples with low classification confidence tend to be ``hard'' samples with low accuracy. Unfortunately, to the best of our knowledge, existing work on the early-exit neural networks has merely assumed this conclusion without verifying it. 
In order to evaluate the relationship between accuracy and confidence, 
a statistical analysis was conducted by us under a typical condition. 
We counted the validation accuracy of the first branch and the backbone within the branch confidence range of [0, 1], 
using a selected category from 100-category classification task in the validation dataset. 
This is visualized by bar charts with reference lines and a table, as shown in Fig.~\ref{fig_8}. 
It should be noted that we use equal sample amounts rather than equal confidence range sizes to count the accuracy. 
Since the distribution of confidence is not uniform, using equal confidence range sizes may cause confidence ranges with small sample amounts to lose statistical significance. 
By sorting the confidence values and dividing them into 15 segments with equal sample amounts, 
we obtained specific confidence range segments as listed in the right table of Fig.~\ref{fig_8}. 
From the upper left section part of Fig. ~\ref{fig_8}, it can be observed that the accuracy of the branch classifier generally tends to increase with the confidence of the branch classifier, 
which is in agreement with the intuition that accuracy increases with sample difficulty. 
In other words, confidence can serve as a reliable criterion for a classifier to evaluate the sample difficulty to some extent. 

\begin{figure}[!ht]
  \centering
  \subfloat[]
  {\includegraphics[width=2.5in]{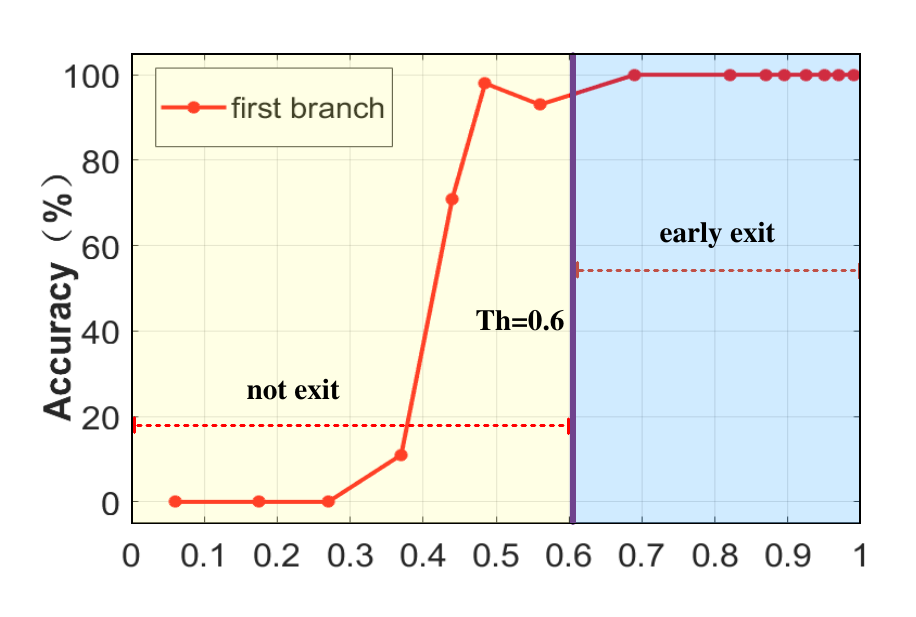}%
  \setlength{\abovecaptionskip}{-10cm}
  \label{fig_9a}}
  \vspace{-5pt}
  \subfloat[]
  {\includegraphics[width=2.5in]{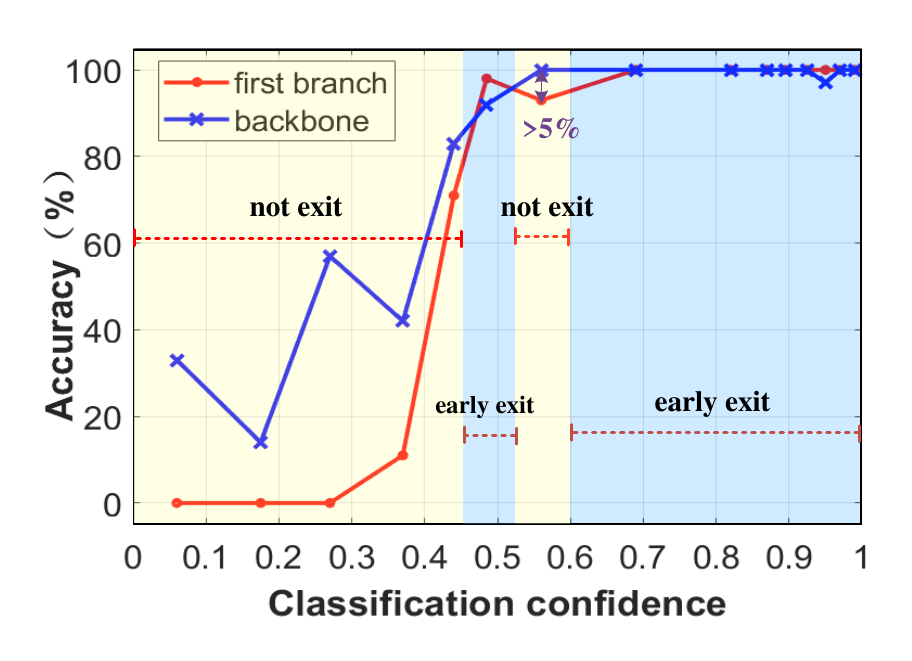}%
  \label{fig_9b}}
  \caption{Analysis diagram for two early-exit strategies (a)conventional early-exit strategy (b)our early-exit strategy.}
  \label{fig_9}
\end{figure}

While there are fluctuations within certain confidence ranges, we believe that this is a normal phenomenon caused by the combined effects of sample distribution and classifier generalization ability. 
For example, the branch classifier accuracy of samples in the 7th segment $((0.52,0.60])$ of the confidence range is lower than that of samples in the 6th segment $((0.45,0.52])$ of the confidence range. 
One possible reason is that the samples in the confidence range $(0.52,0.60]$ are not sufficiently trained and learned by the branch classifier compared to the samples in the confidence range $(0.45,0.52]$, 
resulting in poor accuracy performance of the validation dataset in the actual experiments.

Additionally, the statistical relationship between the branch confidence and the accuracy of the backbone is shown in the lower left of Fig.~\ref{fig_8}. 
As sample difficulty is a relative rather than an absolute concept, classifiers at different locations make different difficulty judgments about the classification confidence. 
Those samples that are considered to have high confidence by the branch classifier are not definitely more accurate under the backbone. 
For instance, the accuracy of the backbone is lower than that of the branch classifier for the samples in the 6th segment $((0.45,0.52])$ and 13th segment $((0.94,0.96])$ of the confidence range. 
This indicates that some samples are more suitable for classification by branch classifier with shallow features, while deep features may adversely affect the classification accuracy of these samples. 
Similarly, those samples that are considered to have low confidence by the branch classifier are not definitely worse under the backbone. 
For instance, the accuracy of the backbone is higher than that of the branch classifier for the samples in the 1st segment $([0,0.12])$ and 2nd segment $((0.12,0.23])$ of the confidence range. 
Therefore, these low-confidence samples are more suitable for classification using backbone to further ensure high accuracy.

\begin{figure*}[!ht]
  \centering
  \includegraphics[width=2.0\columnwidth]{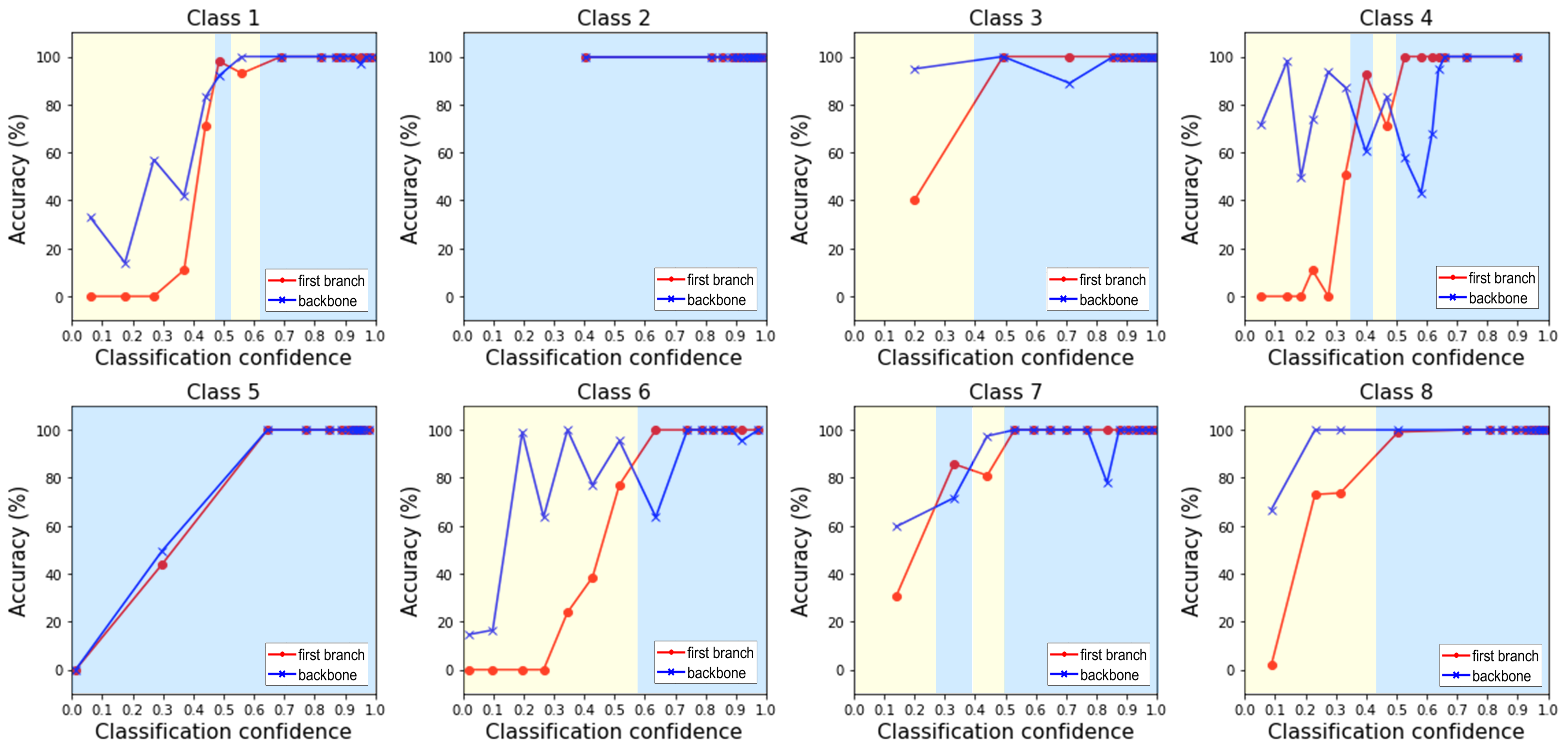}
  \caption{Analysis diagram of our early-exit strategy for different categories under the first branch.}
  \label{fig_10}
\end{figure*}

\subsection{Impact of Early-exit Strategies }

Using the statistical method in the previous section, we keep the reference lines on the bar chats and further compare the differences between the conventional early-exit strategy and our early-exit strategy.
An example of the analysis diagram for two early-exit strategies is given in Fig.~\ref{fig_9}. 
As previously mentioned in Section III, the conventional early-exit strategy compares the confidence with a preset threshold. 
Typically, the threshold is manually set and suggested to be between 0.5 and 0.7. To ensure the accuracy of branch during early-exit, we set a threshold of 0.6 for conventional early-exit strategy in the example of Fig. ~\ref{fig_9}(a), 
where the threshold $Th$ is indicated by the purple line. When the confidence exceeds 0.6, early-exit is performed and marked as the yellow part. 
Otherwise, early-exit cannot be performed and is marked as the blue part. As for our early-exit strategy, 
its analysis diagram is shown in Fig.~\ref{fig_9}(b). Different from the conventional early-exit strategy that uses a manually set threshold as the early-exit condition, 
our early-exit strategy automatically generates early-exit ranges as the early-exit conditions based on the accuracy comparison between branch and backbone. 
If the accuracy of branch is not lower than that of backbone by a certain tolerance $T$, 
or even higher than that of backbone, we consider the current sample can exit early to reduce the computational cost.
If not, we consider the current sample cannot exit early to ensure accuracy. 
In this example, the accuracy tolerance $T$ is marked as a purple line and set at 5\%. Early-exit and no early-exit are still marked as yellow and blue parts, respectively.

Since our early-exit strategy utilizes the accuracy within the per-segment confidence range for comparison, 
it can automatically obtain a more precise early-exit range based on the per-segment endpoints instead of manually setting a single threshold based on empirical experience.
According to the analysis diagram, the difference of the output results is clearly visualized between the two early-exit strategies. 
It can be observed from this example that our early-exit strategy has a 60.0\% (9/15) early-exit rate under the validation set, 
which is 6.7\% higher than the 53.3\% (8/15) early-exit rate of the conventional early-exit strategy. 
The early-exit rate here denotes the percentage of samples that exit at the branch. 
Besides, our early-exit strategy automatically finds classifiers with high accuracy within the confidence range segment, 
especially using branch classifier with higher accuracy within the 6th confidence range segment, which helps us to have an overall accuracy advantage over the conventional early-exit strategy. 
Furthermore, on the difference with the decision-making approach of conventional early-exit strategy, 
our early-exit strategy incorporates multi-category and multi-branch for early-exit.

For multi-category early-exit, we select 8 categories from 100-category classification task and use the first branch to analyze the characteristics of early-exit for different categories, 
as shown in Fig.~\ref{fig_10}. It is evident that each category has its own distinct characteristics, 
as reflected in their sample distribution. Taking class 2 or 5 as an example, the branch and backbone have almost the same sample distribution and trend. 
\begin{figure}[!ht]
  \centering
  \includegraphics[width=1.0\columnwidth]{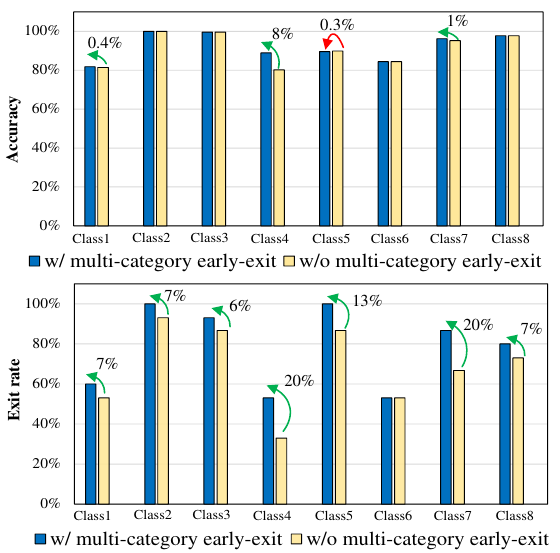}
  \caption{Performance results for each category with and without multi-category early-exit.}
  \label{fig_11}
\end{figure}
\begin{figure}[!ht]
  \vspace{-3mm}
  \centering
  \includegraphics[width=1.0\columnwidth]{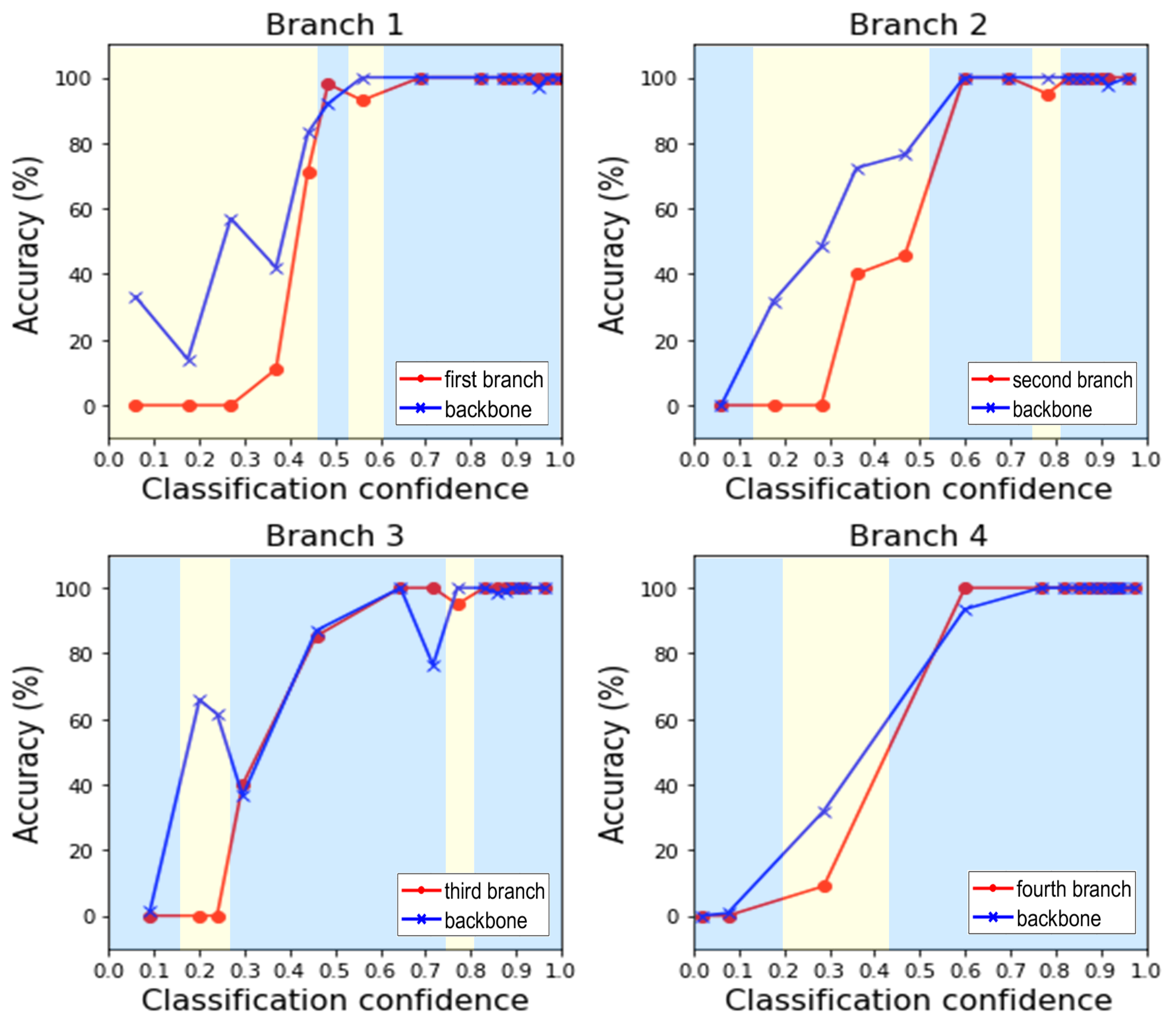}
  \caption{Analysis diagram of our early-exit strategy for different branches under the first category.}
  \label{fig_12}
\end{figure}
This means that the remaining backbone neural network inference at the branch location contributes little to the accuracy of class 2 or 5, 
and thus the samples of these two categories can exit directly at the first branch within the entire confidence range. 
For classes 1, 4, 7 or classes 3, 6, 8, their early-exit ranges have different endpoints and exhibit either discontinuous or continuous characteristics. 
Notably, the continuous early-exit range is similar to that in the conventional early-exit strategy, 
but its location is determined automatically rather than set manually. 
Furthermore, Fig.~\ref{fig_11} gives the specific early-exit rate and accuracy for each category with and without our multi-category early-exit. 
When multi-category early-exit is not used, a uniform threshold of 0.6 is used for all categories to perform the conventional early-exit strategy. 
From the figure, the accuracy for branch and backbone ranges from 81.4\% to 100\% and the exit rate ranges from 33\% to 93\%. 
When our multi-category early-exit is used, there is a maximum 8\% increase in accuracy and a maximum 20\% increase in exit rate for one category. 
Except for a slight decrease in accuracy observed in class 5, other categories show varying degrees of improvement in both accuracy and exit rate or stay unchanged.

\begin{figure}[!ht]
  \centering
  \includegraphics[width=1.0\columnwidth]{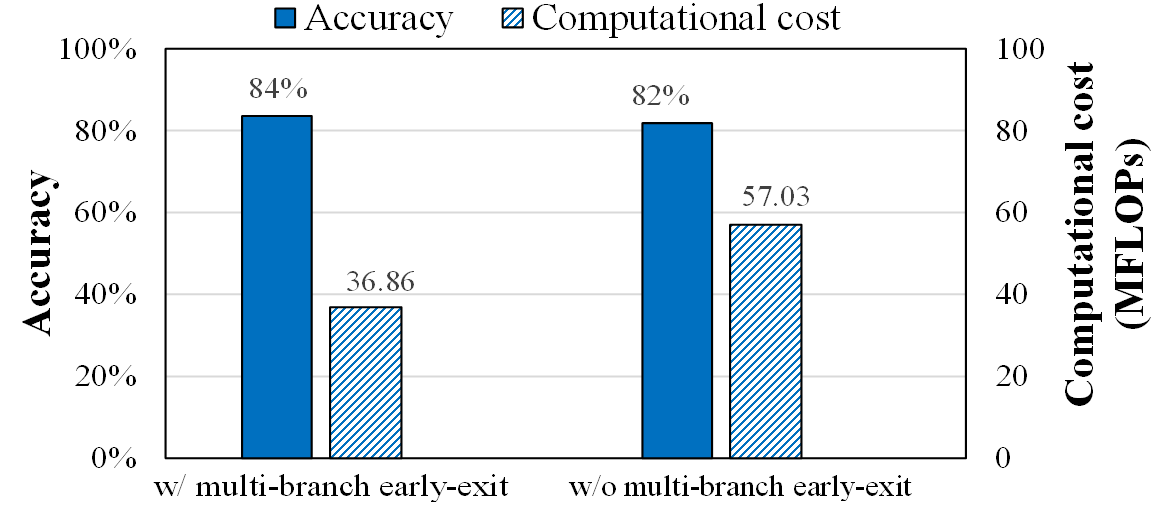}
  \caption{Performance results with and without multi-branch early-exit.}
  \label{fig_13}
\end{figure}
\begin{figure}[!ht]
  \centering
  \includegraphics[width=1.0\columnwidth]{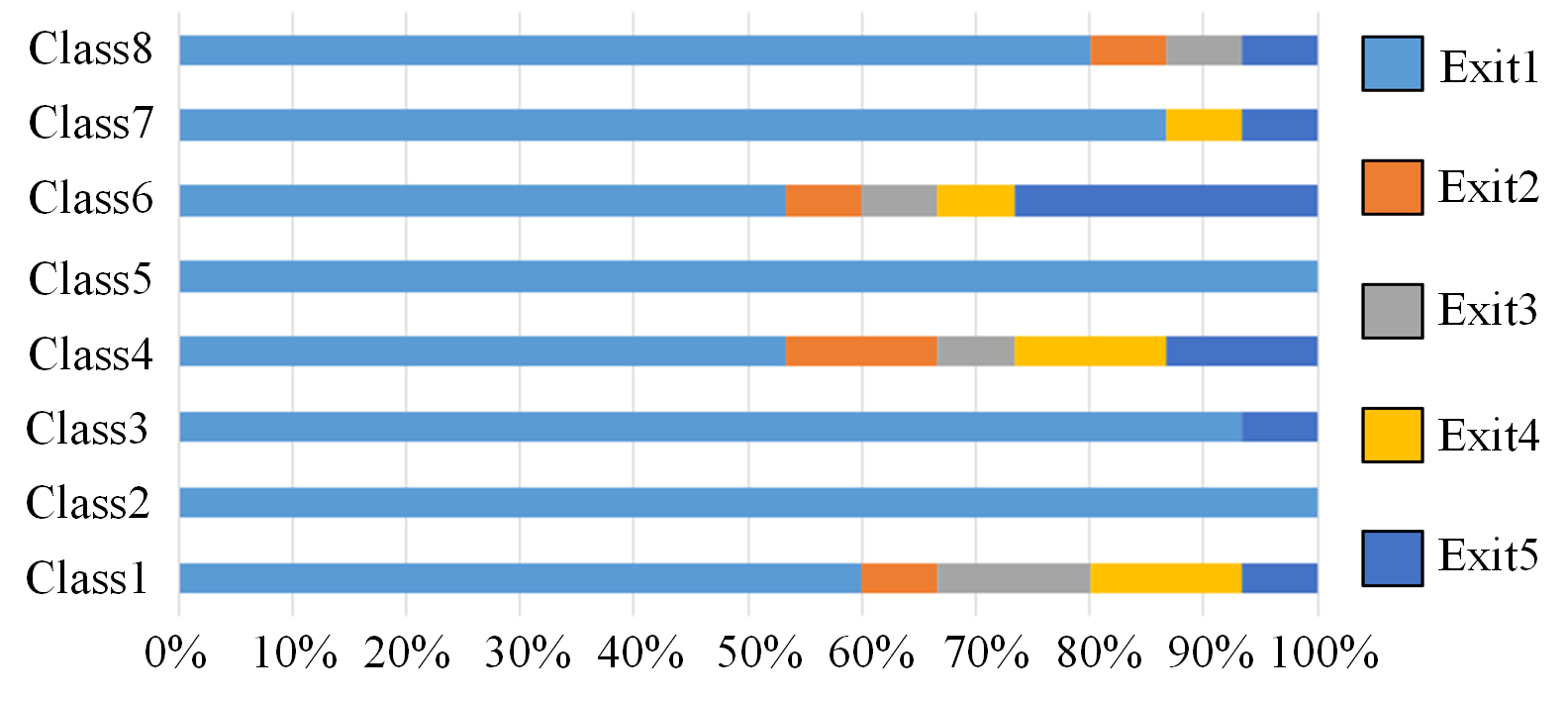}
  \caption{Stacked percentage bar plot of sample size at different exit points for different categories.}
  \label{fig_14}
\end{figure}

For multi-branch early-exit, we choose the first category and use four branches to analyze the characteristics of early-exit for different branches, 
as shown in Fig.~\ref{fig_12}.  These four branches correspond in sequence to the four branches in our hybrid CVNN-RF classifier. 
To facilitate the analysis and experiment, all the validation data goes through all branches instead of exiting early at only one of them. 
It can be seen that each branch has the opportunity to perform early-exit and deep branches have greater exit rates than shallow branches. 
In addition, as the branch position deepens, some previously low-confidence samples move towards the high-confidence regions and the accuracy of the branch gets closer to that of the backbone.
It is concluded that while the shallow branches save a large computational cost, the deep branches still have the opportunity to exit early and guarantee high accuracy. 
Therefore, combining multiple branches with different locations to perform early-exit has a higher theoretical benefit than using a single branch. 
Furthermore, Fig.~\ref{fig_13} gives the specific overall accuracy and computational cost with and without our multi-branch early-exit. 
In the case of not using multi-branch early exit, only the first branch is used. As can be seen from the figure, using multi-branch early-exit has 84\% accuracy and 37.56 million floating point operations (MFLOPs) computational cost, which makes a 2\% accuracy improvement and a 34\% reduction in computational cost over not using multi-branch early-exit.

To further analyze the effect of multi-category early-exit and multi-branch early-exit, a stacked percentage bar graph of the sample size is shown in Fig. 14. Exit points $Exit1$ to $Exit4$ in the figure correspond sequentially to the four early-exit branches, and exit point $Exit5$ is the final exit of the backbone network. It can be seen that there are opportunities for samples to exit at different exit points, while different categories have different proportion of exit samples.
\begin{table}[!ht]
  \caption{\footnotesize{The computational cost of different branch classifiers\label{tab:table4}}}
  \centering
  \setlength{\tabcolsep}{20.5pt}
  \renewcommand{\arraystretch}{2.0}
  \scalebox{1.0}{
  \begin{tabular} {c c}
  \hline
  \hline
  \bf{Branch classifier}	& \bf{Computational cost (MFLOPs)}\\
  \hline
  Random forest 	& 2.05, 1.02, 0.72, 0.51	\\
  \hline
  CONV (3×3)	 & 13.12, 3.28, 2.82, 2.59\\
  \hline
  FC	        &13.11, 3.28, 1.64, 0.41\\
  \hline
  \hline
  \end{tabular}
  }
\end{table}
\begin{figure}[!ht]
  \centering
  \includegraphics[width=1.0\columnwidth]{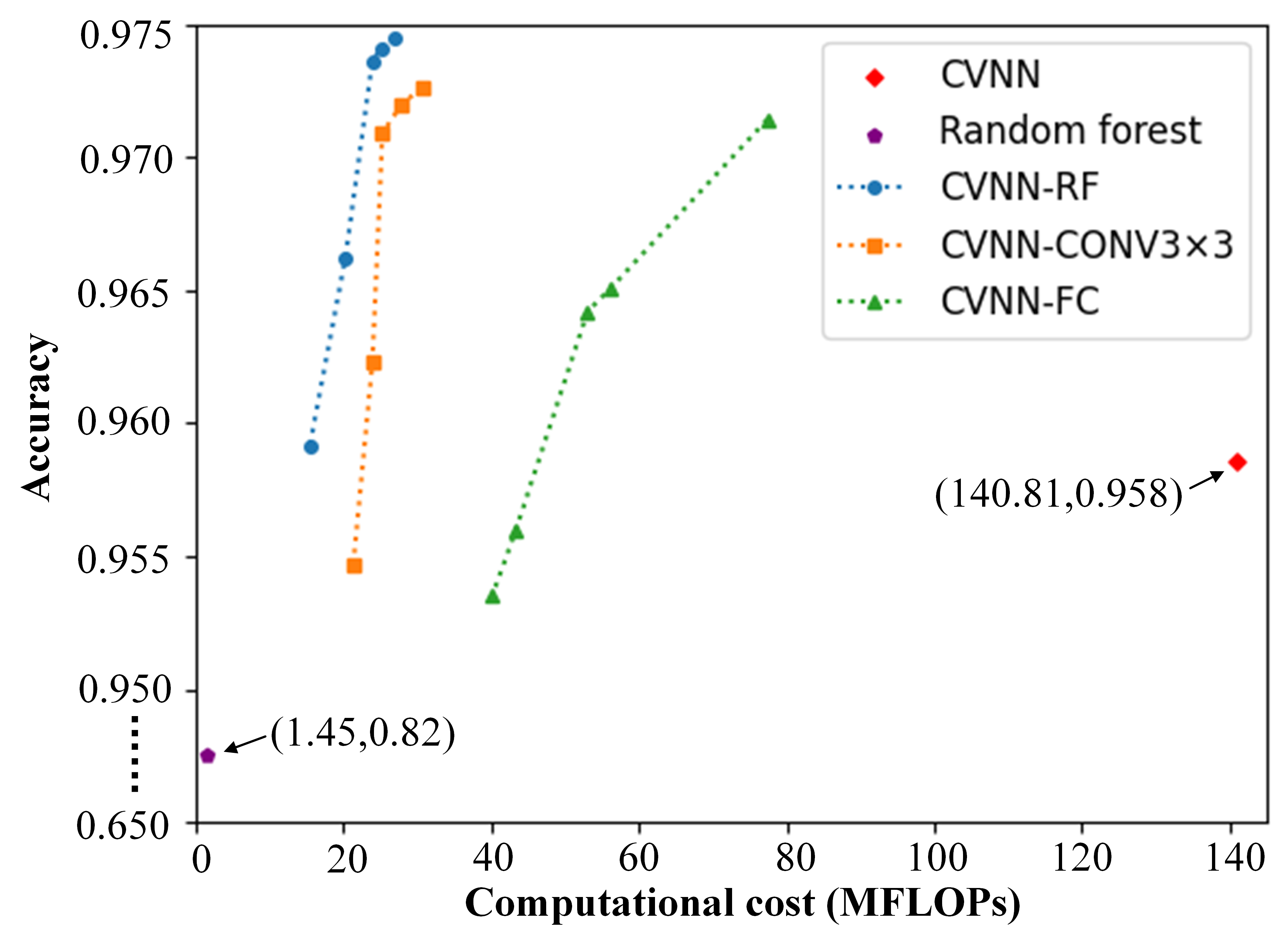}
  \caption{Performance results of different classifiers.}
  \label{fig_15}
\end{figure}
In this case, the first exit point ($Exit1$) exits the most samples, while the exit samples at the remaining exit points vary according to the specific category. For example, all samples from categories 2 and 5 only need to exit at the first exit point, while samples from categories 1 and 4 are distributed across each exit point.

\subsection{Impact of Branch Classifiers Type and Tolerance}
To further investigate the effect of branching classifier types, 
we add CONV branches and FC branches to compare with random forest branches, 
all of which are listed in Table~\ref{tab:table4}. 
If the sample cannot exit early, early-exit neural network has the additional overhead of implementing the branch classifiers. 
Thus, the designed branch classifiers cannot be too complex and need high efficiency. 
In our experiments, four random forest branches are composed of 400 decision trees with a computational cost of 2.05MFLOPs, 1.02MFLOPs, 0.72MFLOPs, and 0.51MFLOPs, 
respectively. For four CONV branches at the early-exit layers, they consist of two 3×3 convolution layers, 
one global pooling layer and one softmax layer, whose computational costs are 13.12MFLOPs, 3.28MFLOPs, 2.82MFLOPs and 2.59MFLOPs respectively. 
Besides, four FC branches are composed of one FC layer and their corresponding computation costs are 13.11MFLOPs, 3.28MFLOPs, 1.64MFLOPs, and 0.41MFLOPs.

Using the aforementioned three branch classifiers, we implemented CVNN-RF, CVNN-CONV3×3, and CVNN-FC for further experiments. As a reliable standard to compare the performance of different classifiers, we set the original CVNN, which is the backbone of CVNN-RF, as the baseline. In addition, using only one random forest classifier as the main classifier was added as another reference point. Fig.~\ref{fig_15} shows the performance results of different classifiers. It is worth noting that the reported computational cost is the average of the computational cost for all test samples with different difficulties. Overall, the performance of the three classifiers is better than the baseline, with CVNN-RF outperforming the other classifiers. While using only one random forest classifier has the lowest computational cost, there is a large accuracy gap compared to the other classifiers, which is unsatisfactory in real-world scenarios. Additionally, there is a trade-off between test accuracy and computational cost based on tolerance. In this case, when the tolerance is low, the classifier can achieve high accuracy but the reduction of the computational cost is limited. On the contrary, the classifier can save more computational cost but sacrifice accuracy. Based on the above analysis, we suggest that the best tolerance can be obtained at the knee point (tolerance=5\%).

\begin{table}[!t]
  \caption{\footnotesize{The performance results of the different branch classifiers (tolerance=5\%)\label{tab:table5}}}
  \centering
  \setlength{\tabcolsep}{3pt}
  \renewcommand{\arraystretch}{2.8}
  \scalebox{1.0}{
  \begin{tabular}{c c c c c}
  \hline
  \hline
  \bf{Class Num} & \bf{Classifier} & \bf{Accuracy (\%)} & \bf{\makecell{Computational \\ cost (MFLOPs)}} & \bf{Exit (\%)}\\
  \hline
  \multirow{4}*{100} & CVNN & 	95.76 & 140.81 & - \\
  \cline{2-5}
  ~ & \makecell{CVNN-\\CONV3×3}	& 97.09 & 25.26 &\makecell{90.09\\(76.49, 3.98,\\4.45,5.17)}\\
  \cline{2-5}
  ~ & CVNN-FC	& 96.42 & 53.05 & \makecell{68.13\\(58.23, 2.97,\\ 3.34,3.59)}\\
  \cline{2-5}
  ~ &CVNN-RF &	97.36 & 23.89 & \makecell{91.05\\(77.57, 4.03,\\ 4.07,5.38)}\\
  \hline
  \multirow{4}*{50} & CVNN & 	96.32 & 140.75 & - \\
  \cline{2-5}
  ~ & \makecell{CVNN-\\CONV3×3}	& 97.41 & 25.11 &\makecell{90.18\\(76.52, 4.01,\\4.38,5.19)}\\
  \cline{2-5}
  ~ & CVNN-FC	& 96.63 & 52.68 & \makecell{68.75\\(58.12, 3.31,\\3.44,3.88)}\\
  \cline{2-5}
  ~ &CVNN-RF &	97.68 & 23.66 & \makecell{91.33\\(77.62, 4.07,\\ 4.23,5.41)}\\
  \hline
  \multirow{4}*{10} & CVNN & 	99.11 & 140.71 & - \\
  \cline{2-5}
  ~ & \makecell{CVNN-\\CONV3×3}	& 99.43 & 23.30 &\makecell{92.14\\(77.52, 4.03,\\4.86,5.73)}\\
  \cline{2-5}
  ~ & CVNN-FC	& 99.12 & 50.56 & \makecell{70.83\\(59.29, 3.31,\\3.94,4.29)}\\
  \cline{2-5}
  ~ &CVNN-RF &	99.56 & 22.08 & \makecell{93.21\\(78.21, 4.10,\\5.10,5.80)}\\
  \hline
  \hline
  \end{tabular}
  }
\end{table}

\begin{figure}[!ht]
  \centering
  \includegraphics[width=1.0\columnwidth]{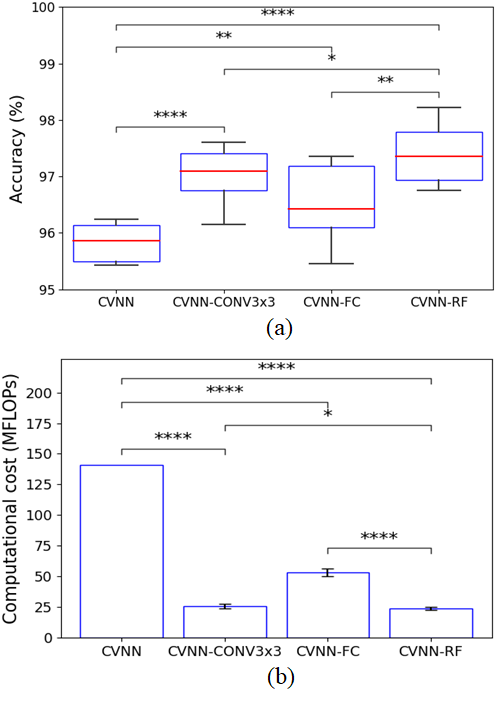}
  \caption{Reliability analysis under 15 Monte Carlo experiments (* denotes significant at the level of 5\%, ** denotes significant at the level of 1\%, **** denotes significant at the level of 0.01\%) (a) box plots for accuracy (b) bar plots for computational cost.}
  \label{fig_16}
\end{figure}

\begin{figure}[!ht]
  \centering
  \includegraphics[width=1.0\columnwidth]{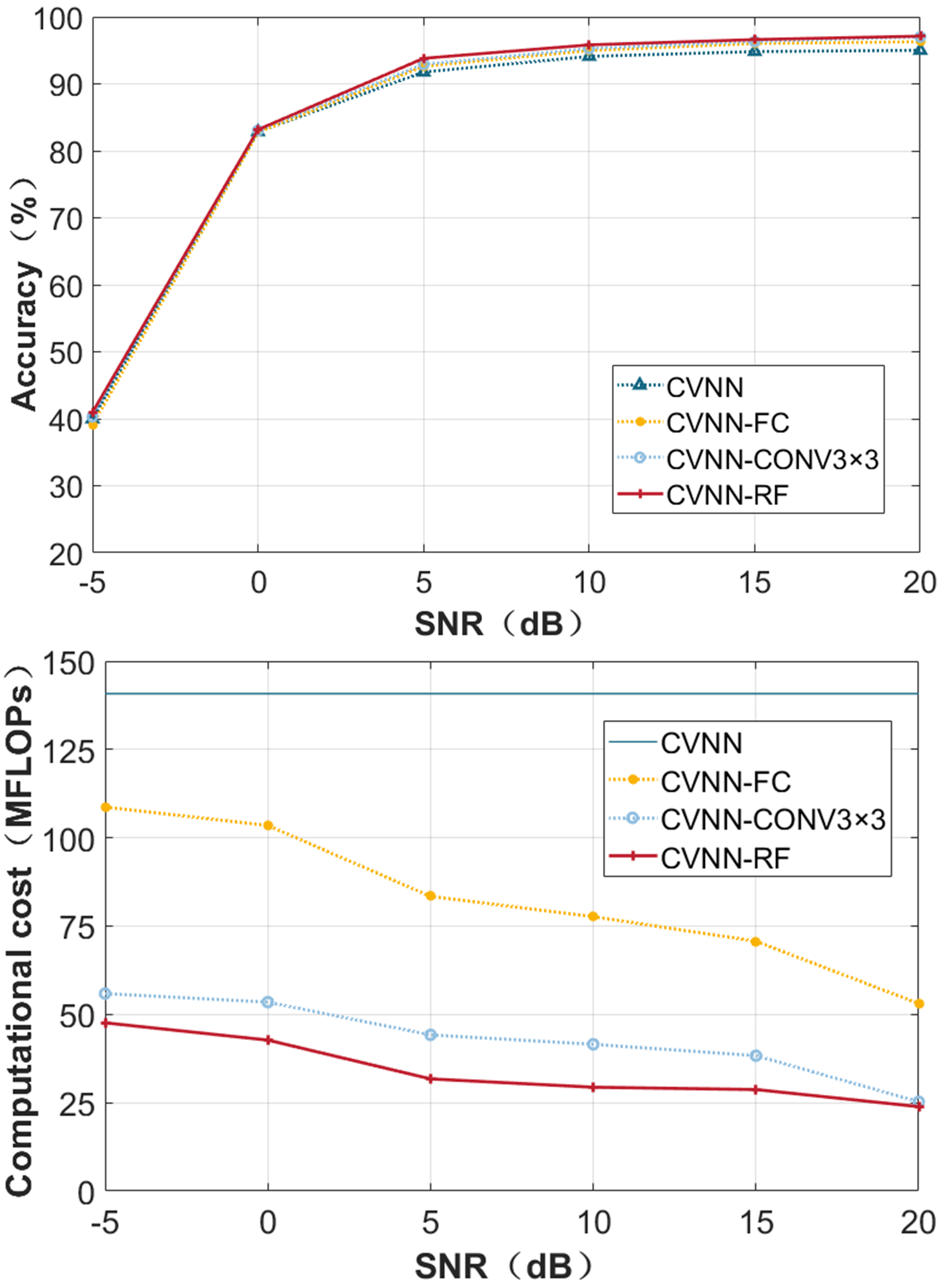}
  \caption{The identification accuracy and computational complexity of our algorithm under different SNR levels.}
  \label{fig_17}
\end{figure}

Table~\ref{tab:table5} highlights the specific accuracy, 
computational cost, and early-exit rate when tolerance is equal to 5\%. 
The “Exit” column denotes the early-exit rate, including the overall early-exit rate and the early-exit rate of each branch. 
From the table, it can be seen that the early-exit neural network is better than the original backbone in terms of accuracy and computational cost, 
especially our proposed CVNN-RF obtains the highest accuracy and lowest computational cost. 
Taking a classification task with 100 categories as an example, 
our CVNN-RF has an accuracy of 97.36\%, which is 1.6\%, 0.27\%, and 0.94\% higher than that of the original CVNN, 
CVNN-CONV3×3, and CVNN-FC. With respect to the computational cost, our CVNN-RF also has the lowest computational cost of 23.89 MFLOPs since CVNN-RF has the highest early-exit rate of 91.05\%. 
Compared to the original CVNN, the computational cost is reduced by 83\%. As the number of classification categories decreases, there is a slight increase in the early-exit rate of the three early-exit neural networks. 
One possible reason is that the difference in accuracy between the branch classifier and the backbone becomes smaller as the classification difficulty decreases. 
As a result, the samples have more chances to exit under the judgment of our early-exit strategy. 
Additionally, the classification difficulty of each category can also influence this phenomenon.

\subsection{Robustness Analysis}
Based on the four models with 5\% tolerance in the previous section, we analyze the accuracy and computational cost statistical results of these models for the 100-category classification task under 15 experiments to verify the reliability of the proposed method. Here, Wilcoxon-based non-parametric tests are used for significance testing to measure the statistical differences between the performance of the different models. Fig. 16(a) illustrates the box plots of the accuracy under the four tested models. For each box, the value of the red horizontal line is the median; the upper and lower two black horizontal lines indicate the maximum and minimum values, respectively; the upper and lower two horizontal lines of the blue box indicate the upper and lower quartiles, respectively. It can be seen that CVNN-CONV3×3, CVNN-FC, and CVNN-RF with our early-exit strategy have a certain degree of accuracy improvement compared with the CVNN baseline, and the significant difference levels are 0.01\%, 1\%, and 0.01\%, respectively. Among the three early-exit models, CVNN-RF has the best accuracy performance, with significant difference levels of 5\% and 1\% compared to CVNN-CONV3×3 and CVNN-FC. Fig. 16(b) illustrates the bar plots of the computational cost under the four tested models. The blue bar reflects the mean value of the computational cost, and the black error bar reflects the distribution of the computational cost. Notably, unlike CVNN which has a fixed computational cost for data computation, the other three models process the samples dynamically through our early-exit strategy, and thus their computational costs are affected by data distribution under different random seeds. It can be seen that CVNN-CONV3×3, CVNN-FC, and CVNN-RF with our early-exit strategy have significant computational cost reduction compared to the CVNN baseline, and the significant difference levels are all 0.01\%. Among the three early-exit models, CVNN-RF has the lowest computational cost, with significant difference levels of 5\% and 0.01\% compared to CVNN-CONV3×3 and CVNN-FC.

To assess the effect of noise on our algorithm, we added noise with six intensities to the original ADS-B dataset. The signal-to-noise ratio (SNR) of the noise-added RF signals takes 5dB as the step and varies from 20dB to -5dB. Under all SNR conditions, We tested our algorithm using the four models trained with the original dataset and early-exit ranges obtained at a tolerance of 5\%. Fig. 17 shows the recognition accuracy and computational complexity of our algorithm at different SNR conditions. It can be seen that our CVNN-RF has the highest accuracy and lowest computational complexity for all SNR conditions. As the SNR decreases, the accuracy of all four models decreases, which is in line with the expected trend. It is worth noting that the SNR affects the computational cost performance of the early-exit networks to some extent. The computational complexity of the 3 early-exit networks increases with decreasing SNR. One possible reason is that the increased noise level makes the samples harder to classify, thereby more samples passing through deeper network layers for processing.

\section{Conclusions and Discussions}
This paper introduces the hybrid CVNN-RF classifier which combines a CVNN backbone with multiple random forest branches. 
Under the scheduling control of our multi-dimensional early-exit strategy, 
the hybrid CVNN-RF is able to dynamically handle RF samples of various difficulties. Specifically, 
for ``easy'' samples, the hybrid model allows them to exit early and obtain classification results from the random forest branch to reduce the computational cost. 
On the other hand, for ``hard'' samples, more layers of the backbone neural network are used for inference to ensure overall accuracy. 
At last, experimental results for a 100-category classification task demonstrate that our hybrid CVNN-RF classifier can achieve a 1.6\% increase in overall accuracy while reducing computational costs by 83\% compared to the original CVNN.

While our method has shown good cost-effectiveness, it is important to recognize its limitations and challenges. First, the low SNR will affect the cost-effectiveness of early-exit models to some extent. Second, deploying early exit models in real-world environments faces additional online challenges, such as hardware resource limitations, small online sample counts, and new data adaptation. 

Regarding the above existing issues, our future research will focus on improving the performance of our early-exit model and its robustness against signal degradation. Due to the plug-and-play nature of the branch module, we see the potential of our approach to adapt to different application requirements of IoT and to be incorporated with other state-of-the-art techniques (e.g., attention mechanisms, data augmentation, model pruning, and neural network architecture search). For the practical application of the IoT, we will further extensively validate the performance of our early-exit model on WiFi, Bluetooth, LTE and other RF data, and explore online deployment schemes using cloud-edge collaboration, transfer learning, open-set identification, and incremental learning.
\section{Acknowledgement}
We would like to sincerely thank Prof. Jun Zhou and Mrs. Linlin Song for their selfless contributions to this research. Their advice and concern helped us overcome many difficulties. We are also grateful to all participants who generously provided their time and support for this study.

\vspace{11pt}
\begin{IEEEbiography}[{\includegraphics[width=1in,height=1.25in,clip,keepaspectratio]{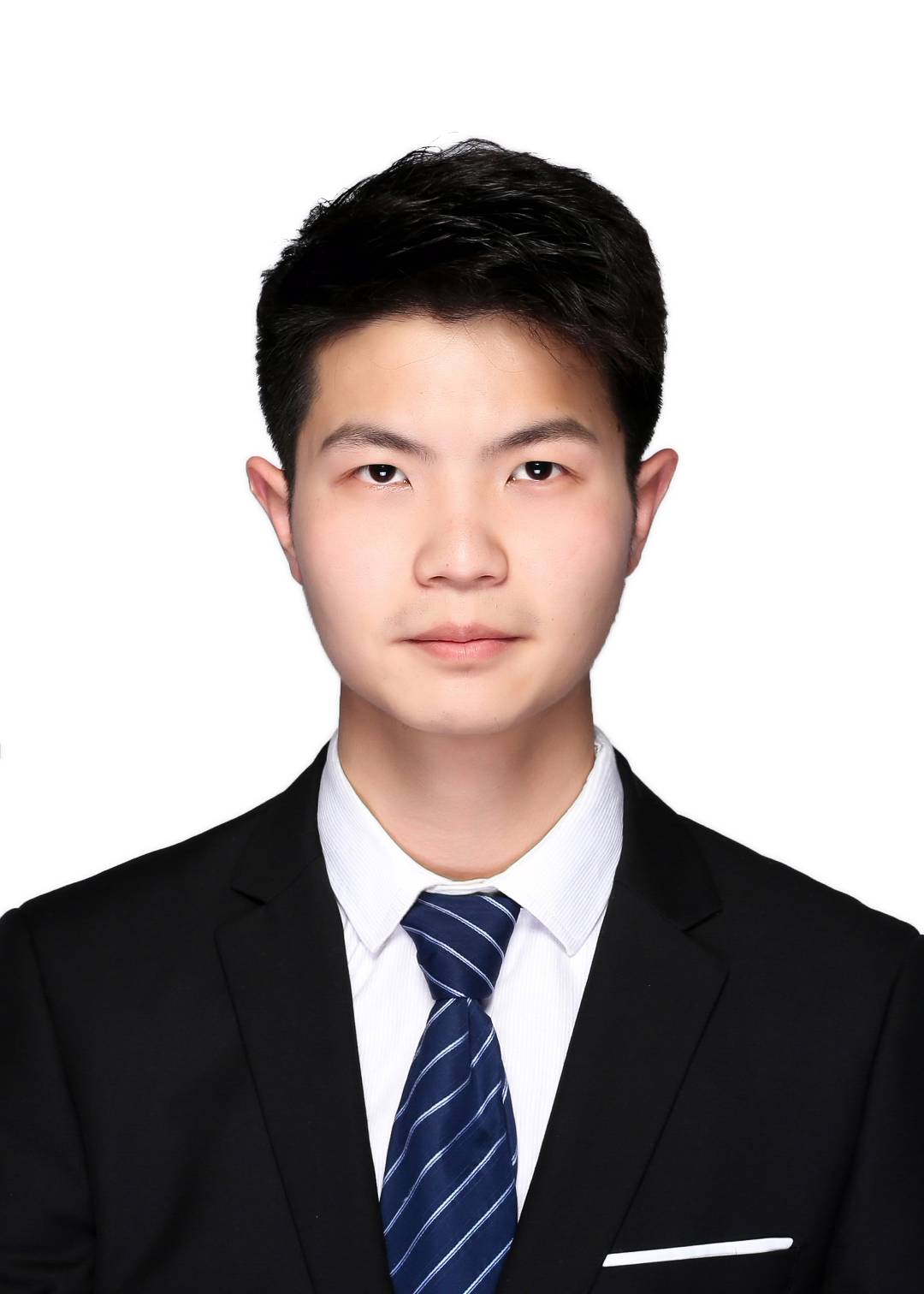}}]{Jiayan Gan}
received the B.Eng. degree from University of Electronic Science and Technology of China (UESTC) in 2013. He is currently pursuing his Ph.D. degree in School of Information and Communication Engineering at UESTC. His research interests include hardware security, specific emitter identification and machine learning processor design.
\end{IEEEbiography}

\vspace{11pt}
\begin{IEEEbiography}[{\includegraphics[width=1in,height=1.25in,clip,keepaspectratio]{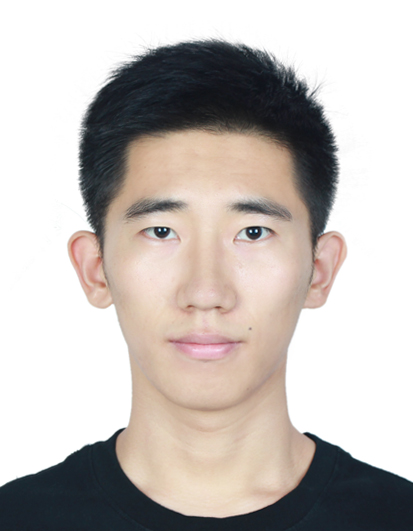}}]{Zhixing Du}
received the M.Eng. degree from University of Electronic Science and Technology of China (UESTC) in 2024. His research interests include hardware acceleration and machine learning algorithms realization in hardware.
\end{IEEEbiography}

\vspace{11pt}
\begin{IEEEbiography}[{\includegraphics[width=1in,height=1.25in,clip,keepaspectratio]{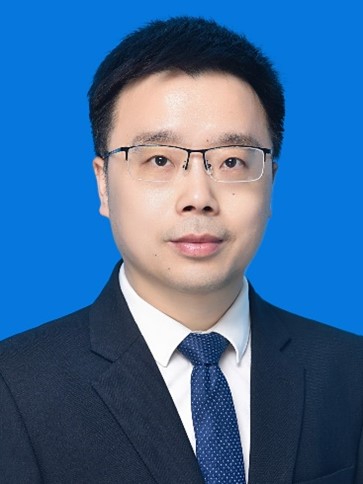}}]{Qiang Li}
received the B.Eng. and M.Phil. degrees in communication and information engineering from the University of Electronic Science and Technology of China (UESTC), Chengdu, China, and the Ph.D. degree in electronic engineering from the Chinese University of Hong Kong (CUHK), Hong Kong, in 2005, 2008, and 2012, respectively. He was a Visiting Scholar with the University of Minnesota, and a Research Associate with the Department of Electronic Engineering and the Department of Systems Engineering and Engineering Management, CUHK. Since November 2013, he has been with the School of Information and Communication Engineering, UESTC, where he is currently a Professor. His research interests focus on machine learning and intelligent signal processing in wireless communications. 
\end{IEEEbiography}

\vspace{11pt}
\begin{IEEEbiography}[{\includegraphics[width=1in,height=1.25in,clip,keepaspectratio]{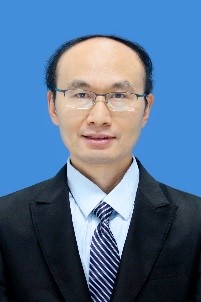}}]{Huaizong Shao}
received the B.S. degree in Optoelectronic Technology from Changchun University of Science and Technology, Changchun, China, in 1992, the M.S. degree in Electrical Engineering from Sichuan University, Chengdu, China, in 1998, and the Ph.D. degree in Information and Communication Engineering from University of Electronic Science and Technology of China (UESTC), Chengdu, China, in 2003. Since 2003, he has been with the School of Information and Communication Engineering, UESTC, where he is currently a Professor. From May 2014 to April 2015, he was a Visiting Scholar with the University of Shefﬁeld, Shefﬁeld, UK. His research interests include communication and radar signal processing. 
\end{IEEEbiography}

\vspace{11pt}
\begin{IEEEbiography}[{\includegraphics[width=1in,height=1.25in,clip,keepaspectratio]{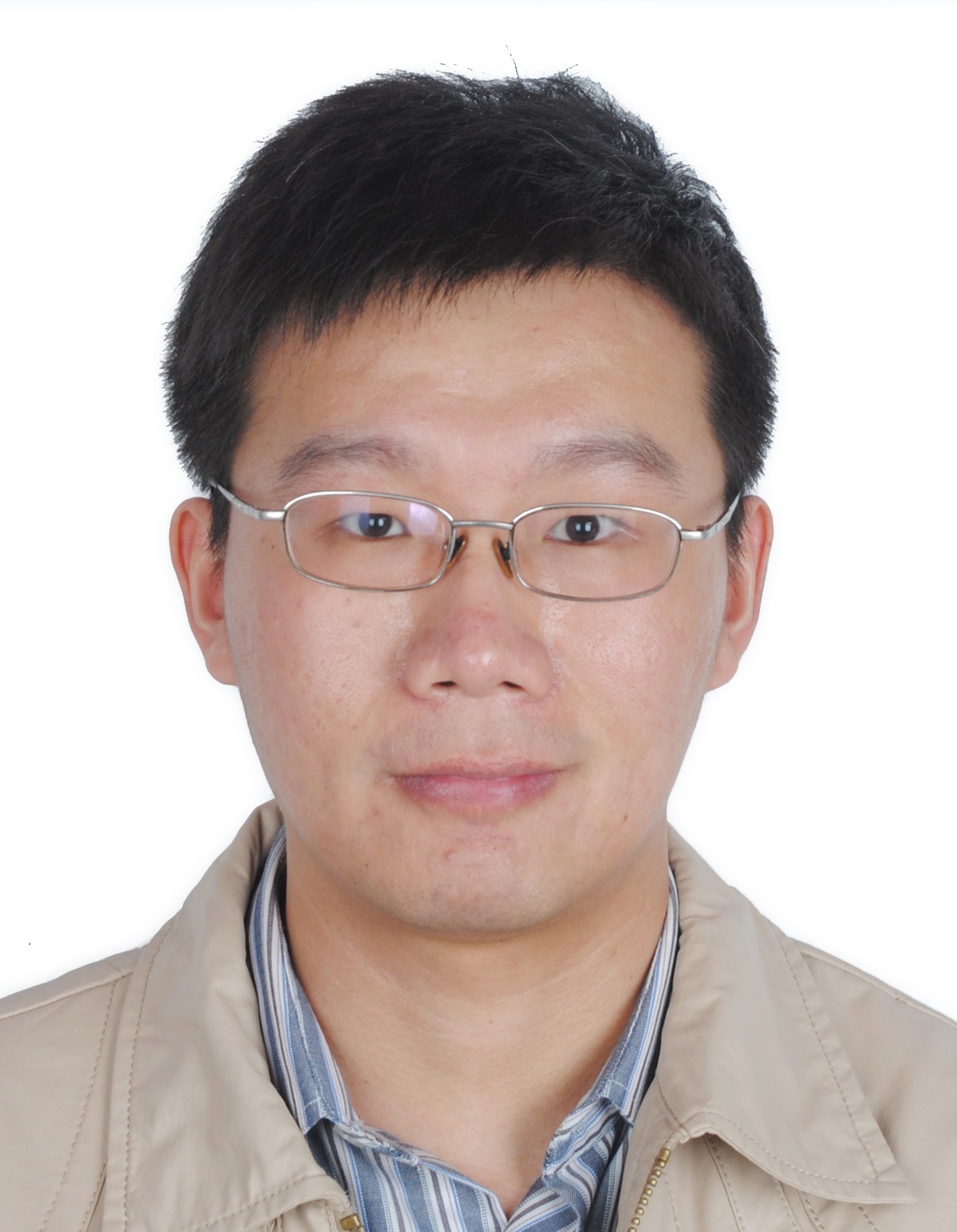}}]{Jingran Lin}
received the B.S. degree in Computer Communication from University of Electronic Science and Technology of China (UESTC), Chengdu, China, in 2001, and the M.S. and Ph.D. degrees in Signal and Information Processing from UESTC in 2005 and 2007, respectively. After his graduation in June 2007, he joined the School of Information and Communication Engineering, UESTC, where he is currently a Full Professor. From January 2012 to January 2013, he was a Visiting Scholar with the University of Minnesota (Twin Cities), Minneapolis, MN, USA. His research interests include the algorithm design and analysis for the intelligent signal processing problems arising from modern communication systems.
\end{IEEEbiography}

\vspace{11pt}
\begin{IEEEbiography}[{\includegraphics[width=1in,height=1.25in,clip,keepaspectratio]{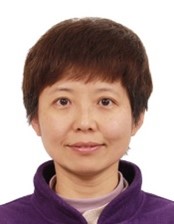}}]{Ye Pan}
received the B.Eng. degree in electronic instrument and measurement and M.Eng. degree in Computer Science from the University of Electronic Science and Technology of China (UESTC), Chengdu, China, in 1992 and 2002, respectively. From Oct. 2008 to Oct. 2009, she was a Visiting Scholar with the University of Southern California (USC), USA. Since April 2002, she has been with the School of Information and Communication Engineering, UESTC, where she is currently an Associate Professor. Her recent research interests focus on machine learning and intelligent signal processing. 
\end{IEEEbiography}

\vspace{11pt}
\begin{IEEEbiography}[{\includegraphics[width=1in,height=1.25in,clip,keepaspectratio]{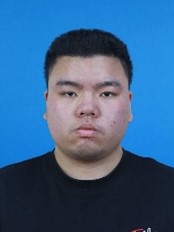}}]{Zhongyi Wen}
received the B.Eng. degree from University of Electronic Science and Technology of China (UESTC) in 2023. He is currently pursuing his Ph.D. degree in School of Information and Communication Engineering at UESTC. His research interests include machine learning, optimization, and adaptive signal processing in the fields of radio frequency fingerprinting identification, AI communications, and medical images. 
\end{IEEEbiography}

\vspace{11pt}
\begin{IEEEbiographynophoto}{Shafei Wang}
received the master's degree in signal and information processing from Beijing Institute of Technology (BIT), Beijing, China, in 1991. He is with the Laboratory of Electromagnetic Space Cognition and Intelligent Control, Beijing, China. His research interest includes signal processing.
\end{IEEEbiographynophoto}

\vfill


\begin{thebibliography}{1}
\bibliographystyle{IEEEtran}

\bibitem{ref1}
Z. Wen, J. Gan, Z. Du, Q. Li, Y. Pan, and H. Shao, “A Hybrid CNN-RF Classifier with Multi- Dimensional Early-Exit Strategy for Radio Frequency Fingerprinting,” in \textit{2023 IEEE International Conference on Communications (ICC)}, Rome, Italy, May. 2023, pp. 2221-2226.

\bibitem{ref2}
A. Ahmed, S. U. Din, E. Alhanaee and R. Thomas, “State-of-the-art in IoT forensic challenges,” in \textit{2022 8th International Conference on Information Technology Trends (ITT)}, Dubai, United Arab Emirates, May. 2022, pp. 115-118.

\bibitem{ref3}
Q. Xu, R. Zheng, W. Saad, and Z. Han, “Device Fingerprinting in Wireless Networks: Challenges and Opportunities,” \textit{IEEE Communications Surveys \& Tutorials}, vol. 18, no. 1, pp. 94–104, 2016.

\bibitem{ref4}
O. Ureten and N. Serinken, “Wireless security through RF fingerprinting,” \textit{Canadian Journal of Electrical and Computer Engineering}, vol. 32, no. 1, pp. 27–33, 2007.

\bibitem{ref5}
A. Jagannath, J. Jagannath, and P. S. P. V. Kumar, “A comprehensive survey on radio frequency (RF) fingerprinting: Traditional approaches, deep learning, and open challenges,” \textit{Computer Networks}, vol. 219, p. 109455, Dec. 2022.

\bibitem{ref6}
Q. Tian et al., “New Security Mechanisms of High-Reliability IoT Communication Based on Radio Frequency Fingerprint,” \textit{IEEE Internet Things J.}, vol. 6, no. 5, pp. 7980–7987, Oct. 2019.

\bibitem{ref7}
D. Reising, J. Cancelleri, T. D. Loveless, F. Kandah, and A. Skjellum, “Radio Identity Verification-Based IoT Security Using RF-DNA Fingerprints and SVM,” \textit{IEEE Internet Things J.}, vol. 8, no. 10, pp. 8356–8371, May 2021.

\bibitem{ref8}
L. Yang, Q. Li, X. Ren, Y. Fang and S. Wang, “Mitigating Receiver Impact on Radio Frequency Fingerprint Identification via Domain Adaptation,” \textit{IEEE Internet Things J.}, 2024.

\bibitem{ref9}
J. Gan et al., “A Zynq-based Platform with Conditional-reconfigurable Complex-valued Neural Network for Specific Emitter Identification,” \textit{IEEE Trans. Instrum. Meas.}, 2024.

\bibitem{ref10}
A. Elmaghbub and B. Hamdaoui, “Leveraging Hardware-Impaired Out-of-Band Information Through Deep Neural Networks for Robust Wireless Device Classification,” \textit{arXiv}, Apr. 23, 2020. {https://arxiv.org/abs/2004.11126v1} (accessed May 31, 2023).

\bibitem{ref11}
S. Deng, Z. Huang, X. Wang, and G. Huang, “Radio Frequency Fingerprint Extraction Based on Multidimension Permutation Entropy,” \textit{International Journal of Antennas and Propagation}, vol. 2017, pp. 1–6, 2017.

\bibitem{ref12}
H. Patel, “Non-parametric feature generation for RF-fingerprinting on ZigBee devices,” in \textit{2015 IEEE Symposium on Computational Intelligence for Security and Defense Applications (CISDA)}, May 2015, pp. 1–5.

\bibitem{ref13}
S. Ur Rehman, K. Sowerby, and C. Coghill, “RF fingerprint extraction from the energy envelope of an instantaneous transient signal,” in \textit{2012 Australian Communications Theory Workshop (AusCTW)}, Jan. 2012, pp. 90–95.

\bibitem{ref14}
K. Sankhe, M. Belgiovine, F. Zhou, S. Riyaz, S. Ioannidis, and K. Chowdhury, “ORACLE: Optimized Radio clAssification through Convolutional neuraL nEtworks,” in \textit{2019 IEEE Conference on Computer Communications (INFOCOM)}, Paris, France, Apr. 2019, pp. 370–378.

\bibitem{ref15}
K. Sankhe et al., “No Radio Left Behind: Radio Fingerprinting Through Deep Learning of Physical-Layer Hardware Impairments,” \textit{IEEE Trans. Cogn. Commun. Netw.}, vol. 6, no. 1, pp. 165–178, Mar. 2020.

\bibitem{ref16}
L. Zong, C. Xu, and H. Yuan, “A RF Fingerprint Recognition Method Based on Deeply Convolutional Neural Network,” in \textit{2020 IEEE 5th Information Technology and Mechatronics Engineering Conference (ITOEC)}, Chongqing, China, Jun. 2020, pp. 1778–1781.

\bibitem{ref17}
A. Gritsenko, Z. Wang, T. Jian, J. Dy, K. Chowdhury, and S. Ioannidis, “Finding a ‘New’ Needle in the Haystack: Unseen Radio Detection in Large Populations Using Deep Learning,” in \textit{2019 IEEE International Symposium on Dynamic Spectrum Access Networks (DySPAN)}, Newark, NJ, USA, Nov. 2019, pp. 1–10.

\bibitem{ref18}
K. He, X. Zhang, S. Ren, and J. Sun, “Identity Mappings in Deep Residual Networks.” \textit{arXiv}, Jul. 25, 2016. Accessed: Jun. 07, 2023. [Online]. Available: http://arxiv.org/abs/1603.05027.

\bibitem{ref19}
H. Gu, L. Su, W. Zhang and C. Ran, “Attention is Needed for RF Fingerprinting.” in \textit{IEEE Access}, vol. 11, pp. 87316-87329, 2023.

\bibitem{ref20}
J.Yang, H. Gu, C. Hu, X. Zhang, G. Gui, and H. Gacanin, “Deep Complex-Valued Convolutional Neural Network for Drone Recognition Based on RF Fingerprinting.” \textit{Drones}, vol. 6, 2022.

\bibitem{ref21}
S. Mann and S. Haykin, “‘Chirplets’ and ‘warblets’: novel time–frequency methods,” \textit{Electronics Letters}, vol. 28, no. 2, pp. 114–116, Jan. 1992.

\bibitem{ref22}
S. G. K. Patro and K. K. sahu, “Normalization: A Preprocessing Stage,” \textit{International Advanced Research Journal in Science, Engineering and Technology}, pp. 20–22, Mar. 2015.

\bibitem{ref23}
P. Panda, A. Sengupta, and K. Roy, “Conditional Deep Learning for Energy-Efficient and Enhanced Pattern Recognition,” in \textit{Proceedings of the 2016 Design, Automation \& Test in Europe Conference \& Exhibition (DATE)}, 2016, pp. 475–480.

\bibitem{ref24}
S. Zangeneh, S. Pruett, S. Lym, and Y. N. Patt, “BranchNet: A Convolutional Neural Network to Predict Hard-To-Predict Branches,” in \textit{2020 53rd Annual IEEE/ACM International Symposium on Microarchitecture (MICRO)}, Oct. 2020, pp. 118–130.

\bibitem{ref25}
K. Park, C. Oh, and Y. Yi, “BPNet: Branch-pruned Conditional Neural Network for Systematic Time-accuracy Tradeoff,” in \textit{2020 57th ACM/IEEE Design Automation Conference (DAC)}, San Francisco, CA, USA, Jul. 2020, pp. 1–6.

\bibitem{ref26}
C. Trabelsi et al., “Deep Complex Networks,” \textit{arXiv:1705.09792 [cs]}, Feb. 2018, Accessed: Nov. 03, 2021. [Online]. Available: http://arxiv.org/abs/1705.09792.

\bibitem{ref27}
M. Belgiu and L. Drăguţ, “Random forest in remote sensing: A review of applications and future directions,” \textit{ISPRS Journal of Photogrammetry and Remote Sensing}, vol. 114, pp. 24–31, Apr. 2016.

\bibitem{ref28}
Y. Tu et al., “Large-scale real-world radio signal recognition with deep learning,” \textit{Chinese Journal of Aeronautics}, vol. 35, no. 9, pp. 35–48, Sep. 2022.

\bibitem{ref29}
Z. Xu, G. Han, L. Liu, H. Zhu, and J. Peng, “A Lightweight Specific Emitter Identification Model for IIoT Devices Based on Adaptive Broad Learning,” \textit{IEEE Trans. Ind. Inf.}, pp. 1–10, 2022.

\bibitem{ref30}
Y. Wang, J. Yang, M. Liu, and G. Gui, “LightAMC: Lightweight Automatic Modulation Classification via Deep Learning and Compressive Sensing,” \textit{IEEE Trans. Veh. Technol.}, vol. 69, no. 3, pp. 3491–3495, Mar. 2020.

\bibitem{ref31}
J. Yu, A. Hu, G. Li, and L. Peng, “A Robust RF Fingerprinting Approach Using Multisampling Convolutional Neural Network,” \textit{IEEE Internet Things J.}, vol. 6, no. 4, pp. 6786–6799, Aug. 2019.

\end{thebibliography}
\end{document}